\documentclass[a4paper,11pt]{article}
\pdfoutput=1 
\usepackage{jcappub} 
\usepackage{amssymb,amsthm,bbold,bm,mathtools,slashed}
\usepackage{comment,enumerate,footnote,graphicx,subfloat,relsize}
\usepackage{array,tabularx,tabu,multirow,framed,xcolor}
\usepackage[font=small]{caption}
\usepackage{subcaption}
\allowdisplaybreaks
\usepackage{cleveref}
\crefformat{pluralequation}{#2\black{eqs.~(}#1\black{)}#3}
\Crefformat{pluralequation}{#2\black{Equations~(}#1\black{)}#3}
\usepackage{tikz}
\usetikzlibrary{arrows,shapes}
\usetikzlibrary{trees,patterns}
\usetikzlibrary{matrix,arrows} 				
\usetikzlibrary{positioning}				  
\usetikzlibrary{calc,through}				  
\usetikzlibrary{decorations.pathreplacing}  
\usepackage{pgffor}							

\usetikzlibrary{decorations.pathmorphing}	
\usetikzlibrary{decorations.markings}
\tikzset{
	>=stealth', 
    vector/.style={decorate, decoration={snake}, draw},
	provector/.style={decorate, decoration={snake,amplitude=2.5pt}, draw},
	antivector/.style={decorate, decoration={snake,amplitude=-2.5pt}, draw},
	bigvector/.style={decorate, decoration={snake,amplitude=4pt}, draw},
    fermion/.style={draw=black, postaction={decorate},
        decoration={markings,mark=at position .55 with {\arrow[draw=black]{>}}}},
    fermionbar/.style={draw=black, postaction={decorate},
        decoration={markings,mark=at position .55 with {\arrow[draw=black]{<}}}},
    fermionnoarrow/.style={draw=black},
    gluon/.style={decorate, draw=black,
        decoration={coil,amplitude=4pt, segment length=5pt}},
    scalar/.style={dashed,draw=black, postaction={decorate},
        decoration={markings,mark=at position .55 with {\arrow[draw=black]{>}}}},
    scalarbar/.style={dashed,draw=black, postaction={decorate},
        decoration={markings,mark=at position .55 with {\arrow[draw=black]{<}}}},
    scalarnoarrow/.style={dashed,draw=black},
    momentum/.style={draw=black, postaction={decorate},
        decoration={markings,mark=at position 1 with {\arrow[draw=black]{>}}}},
    antimomentum/.style={draw=black, postaction={decorate},
        decoration={markings,mark=at position 0.1 with {\arrow[draw=black]{<}}}}
}

\tikzstyle{block} = [draw, rectangle, 
    minimum height=3em, minimum width=6em]

\title{
Radiation back-reaction during dark-matter freeze-out via metastable bound states
}

\author{Christiana Vasilaki and Kalliopi Petraki}
\affiliation{Laboratoire de Physique de l'École Normale Supérieure, ENS, Université PSL, CNRS, Sorbonne Université, Université Paris Cité, F-75005 Paris, France}

\emailAdd{christiana.vasilaki@phys.ens.fr}
\emailAdd{kalliopi.petraki@phys.ens.fr}

\abstract{
The formation and decay of metastable bound states can deplete significantly the density of multi-TeV thermal-relic dark matter. The effect depends on the interplay of bound-state formation, ionisation, transition and decay processes. Existing calculations take into account bound-state ionisation and excitations due to the radiation of the thermal bath. However, the dynamics of Hydrogen recombination suggests that the resonant radiation produced in bound-state formation or de-excitations may backreact, ionising or exciting the bound states thus impeding recombination. In this paper we examine this effect in the context of dark-matter freeze-out. To this end, we employ the \emph{generalised Saha equilibrium equation for metastable bound states}, and discuss its salient features. We show that, in sharp contrast to Hydrogen recombination, the radiation produced during dark matter freeze-out is more likely to thermalise or redshift, rather than ionise or excite the metastable bound states.
This holds not only for the low-energy (resonant) radiation produced in bound-state formation and transition processes, but also for the high-energy radiation produced in dark-matter annihilations and bound-state decays. While our computations are carried out in a minimal dark $U(1)$ model, our conclusions only strengthen in more complex models.
}


\begin{document}
\maketitle
\flushbottom

\section{Introduction \label{Sec:Introduction}}
One of the most widely studied production scenarios for dark matter (DM) is its thermal decoupling from the primordial plasma. In viable thermal-relic DM scenarios with DM masses around or above the TeV scale, the DM interactions manifest as \emph{long-ranged}, their range being larger than the de Broglie wavelength of the interacting DM particles. This is confirmed by explicit calculations in specific models and supported by model-independent unitarity arguments~\cite{Baldes:2017gzw}. Long-range interactions imply the existence of bound states. Metastable DM bound states can form and decay into radiation, thereby opening additional annihilation channels for DM. This, in turn, may affect the DM freeze-out and alter the predicted mass-coupling relations that account for the observed DM density~\cite{vonHarling:2014kha}. The effect of metastable bound states on the DM depletion in the early universe has been studied in many theories~\cite{vonHarling:2014kha, Ellis:2015vaa, Petraki:2015hla, Petraki:2016cnz, Kim:2016zyy,Biondini:2017ufr,Biondini:2018pwp,Biondini:2018ovz,Biondini:2019int,Binder:2018znk,Ko:2019wxq,Baldes:2017gzw, Harz:2018csl,Harz:2019rro, Oncala:2018bvl, Oncala:2019yvj,Binder:2019erp,Binder:2020efn,Binder_2022,Oncala:2021tkz,Oncala:2021swy,Bottaro:2021snn,Binder:2021vfo,Binder:2023ckj}, and has been shown to affect the DM density even up to orders of magnitude. Metastable bound states contribute also to the expected DM indirect detection signals~\cite{Pospelov:2008zw, MarchRussell:2008tu,An:2016kie,An:2016gad,Asadi:2016ybp,Cirelli:2018iax,Baldes:2017gzu,Baldes:2020hwx}.

The efficacy of metastable bound states in depleting DM depends on the interplay of bound-state formation (BSF), bound-to-bound transition, bound-state decay and their inverse processes. Ionisation (a.k.a.~dissociation) processes in general suppress the net effect. Current calculations take into account the ionisation of bound states due to the radiation of the thermal bath, with the qualitative result being that the DM depletion via BSF becomes most significant when the plasma temperature approaches or drops below the binding energy. 

Recalling the related physics of recombination however, it is well-understood that the formation of Hydrogen in the early universe is impeded not only by the thermal radiation, but also by the backreaction of the resonant photons produced in the capture of proton-electron pairs into the ground state; once produced, these photons ionise very efficiently other bound states, thereby hindering recombination~\cite{Peebles:1968ja}. Consequently, Hydrogen recombination occurs more efficiently via formation of excited states and their subsequent transition to lower-lying states. A multi-level atom approach, considering the evolution of each level and the back-reaction of the resonant radiation including bound-bound and bound-free transitions, is required to accurately calculate cosmological recombination, with the redshifting of the resonant radiation due to the Hubble expansion playing a critical role~\cite{dubrovich2005dynamics,Seager_2000,PhysRevD.82.123502}.

This paradigm raises the question whether the resonant radiation produced in BSF and bound-state de-excitations during the DM thermal decoupling may similarly backreact, ionising or exciting DM bound states, thus affecting the DM density. Higher-energy radiation produced in DM direct annihilations and bound-state decays during freeze-out could also potentially dissociate or excite bound states. The purpose of the present work is to examine the back-reaction of these radiation components on the metastable DM bound states, in order to check the validity of the current calculations of the DM abundance when metastable bound states exist in the spectrum of the theory.

To do so, we consider a minimal DM model in which DM is a pair of Dirac fermions coupled to a massless dark photon. Particle-antiparticle pairs can form dark positronium bound states that subsequently decay into dark photons; the DM depletion via this channel becomes significant if DM is heavier than a few TeV~\cite{vonHarling:2014kha}. Here, we extend the calculation of~\cite{vonHarling:2014kha} to include the first excited bound level; besides improving the accuracy of the calculation, this allows to consider the radiation produced in bound-to-bound transitions in addition to that of BSF, and resembles better the multi-level system that describes Hydrogen recombination.

Considering the effect of metastable bound states on the DM abundance, even without taking into account the backreaction of the radiation produced during freeze-out, requires in principle solving a system of coupled Boltzmann equations for the number densities of the free and bound species~\cite{vonHarling:2014kha}. However, metastable bound states reach and remain in a quasi-steady state during all relevant times, due to their large interaction rates: at high temperatures, BSF and ionisation are very efficient, while at lower temperatures, the bound-state decay rate is typically larger than the Hubble expansion rate. This quasi-steady state allows reducing the coupled equations to a single effective Boltzmann equation for the DM relic density~\cite{Ellis:2015vaa,Binder:2021vfo}, and generalises the well-known Saha equilibrium for stable to metastable bound states~\cite{Binder:2021vfo}. 

Investigating the back-reaction of the radiation produced during DM freeze-out on the bound states requires considering the (un-integrated) Boltzmann equation for the radiation phase-space density. This must include the effects of red-shifting of radiation due to the Hubble expansion, thermalisation of radiation via elastic scattering on ions, ionisation, excitation and inverse decays of bound states, pair creation of free DM particles, as well as the inverse processes. We formulate this equation and then estimate the rates of the various effects using the densities of the free and bound DM species obtained when neglecting any radiation back-reaction. Essential in this step is the use of the generalised Saha equilibrium for the bound-state densities~\cite{Binder:2021vfo}. Comparing the rates of the various effects allows us to robustly conclude that, in contrast to Hydrogen recombination, the resonant and higher-energy radiation produced during DM freeze-out does not affect the DM bound states and the DM density, therefore validating current calculations in this respect.

This paper is organised as follows. In \cref{Sec:FreezeOut}, we introduce the model, summarise the relevant interaction rates, and compute the DM relic density using the effective Boltzmann equation that is based on the generalised Saha equilibrium. We then offer a detailed description of the evolution of the metastable-bound-state densities in this context. 
In \cref{Sec:PhotonDist}, we derive the kinetic equation for the dark photon distribution function, compare the rates of the processes in which the dark photons participate, and discuss the generalisation to other models. We conclude in section \ref{Sec:Conclusions}.

\section{Dark-matter freeze-out \label{Sec:FreezeOut}}

\subsection{The model \label{sec:Freezeout_Model}}

We consider a DM model featuring a dark $U(1)$ gauge symmetry. The DM particles, denoted as $\chi$, are Dirac fermions of mass $m_\chi$ interacting with a massless dark photon $\gamma$.\footnote{We choose the symbol $\gamma$ instead of e.g.~$\gamma_{\mathsmaller{D}}$ to keep the notation simple. Since we will not be referring to the ordinary photons in this work, there is no risk of confusion.} 
The Lagrangian of the model is
\begin{equation}
\mathcal{L}=\bar{\chi}\left(i \slashed{D}-m_\chi\right) \chi-\frac{1}{4} F_{\mu \nu} F^{\mu \nu},
\end{equation}
where $D^\mu=\partial^\mu+i g A^\mu$ is the covariant derivative, $F^{\mu \nu}=\partial^\mu A^\nu-\partial^\nu A^\mu$ is the field strength tensor, $g$ denotes the dark coupling constant, and $\alpha=g^2 / 4 \pi$ corresponds to the dark fine structure constant.

In this model, $\chi\bar{\chi}$ pairs can annihilate directly into dark photons, or form dark positronium bound states that are unstable against decay into dark photons. The dark positronium bound states can be characterised by the quantum numbers ${\cal B} = \{n\ell m_\ell;~s,m_s\}$, where $n\ell m_\ell$ are the standard principal and orbital angular momentum quantum numbers, and $s,m_s$ characterise the total spin of the $\chi\bar{\chi}$ pair, with the `para' and `ortho' configurations corresponding to $s=0$ (singlet) and $s=1$ (triplet). The mass of the bound states is $m_{\cal B} = 2 m_{\chi} + {\cal E}_{\cal B}$, where the binding energy ${\cal E}_{\cal B}<0$ depends at leading order in $\alpha$ only on the principal quantum number $n$: $\mathcal{E}_n=- \mu \alpha^2 / (2 n^2)$, with $\mu \equiv m_\chi/2$ being the reduced mass of the $\chi\bar{\chi}$ pair.

\subsection{Boltzmann equations \label{sec:FreezeOut_Boltzmann}}

The evolution of the densities of the free and bound species is governed by a system of coupled Boltzmann equations that describes the interplay among DM direct annihilation, BSF, bound-state decay, bound-to-bound transitions and the respective inverse processes. Let $Y_{\chi} = n_{\chi}/s$ and $Y_{\cal B} = n_{\cal B}/s$ be the comoving number densities of the free species and the bound states with ${\cal B}$ denoting collectively the quantum numbers of each bound state. As standard, $s = \left(2 \pi^2 /45\right) g_{* S} T^3$ is the entropy density of the universe and $g_{* S}$ the corresponding degrees of freedom (dof). The system of Boltzmann equations describing the evolution of the comoving densities is
\begin{subequations}
\label{eq:SystemBoltzmann}
\begin{align}
\frac{d Y_\chi}{d x} 
=& -\frac{\lambda}{x^2} \left[ 
\langle\sigma^{\rm ann} v_{\rm rel}\rangle 
\left(Y_{\chi}^{2}-\left(Y_{\chi}^{\rm eq}\right)^{2} \right)  
+ \sum_{\cal B}  \langle \sigma_{\cal B}^{\rm BSF} v_{\rm rel}\rangle
\left(Y_{\chi}^{2}-\frac{Y_{\cal B}}{Y_{\cal B}^{\rm eq}} \left(Y_{\chi}^{\rm eq}\right)^{2}\right) 
\right], 
\label{eq:SystemBoltzmann_chi}
\\
\frac{d Y_{\cal B}}{d x} 
=& -\Lambda x \left[ 
  \bar{\Gamma}_{\cal B}^{\rm dec} \left(Y_{\cal B} - Y_{\cal B}^{\rm eq}\right) 
+ \bar{\Gamma}_{\cal B}^{\rm ion} \left(Y_{\cal B} - \left(\frac{Y_{\chi}}{Y_{\chi}^{\rm eq}} \right)^{2} Y_{\cal B}^{\rm eq}\right) \right. \nonumber 
\\
&\left. + \sum_{{\cal B}' \neq {\cal B}} \bar{\Gamma}_{{\cal B} \rightarrow {\cal B}'}^{\rm trans} 
\left(Y_{\cal B}-\frac{Y_{{\cal B}'}}{Y_{{\cal B}'}^{\rm eq}} Y_{\cal B}^{\rm eq}\right) \right],
\end{align}
\end{subequations}
where $x \equiv m_{\chi} / T$ is the standard time variable and
\begin{align}
\lambda & \equiv \sqrt{\frac{\pi}{45}} m_{{\rm Pl}} m_{\chi} g_*^{1 / 2}, 
\label{eq:lambda}
\\
\Lambda & \equiv \frac{\lambda}{s x^3}=\sqrt{\frac{45}{4 \pi^3}} \frac{m_{{\rm Pl}}}{m_{\chi}^2} \frac{g_*^{1 / 2}}{g_{* S}},
\label{eq:Lambda}
\\
g_*^{1/2} &\equiv \frac{g_{* S}}{g_{* \rho}} 
\left(1 - \frac{x}{3 g_{* S}} \frac{d g_{* S}}{dx} \right),
\label{eq:gstar}
\end{align}
with $g_{*\rho}$ being the energy dof of the universe and $m_{\rm Pl} = 1.22 \cdot 10^{19}$~GeV being the Planck mass. The equilibrium densities in the non-relativistic regime are
\begin{align}
Y_{\chi}^{\rm eq} & \simeq \frac{90}{(2 \pi)^{7 / 2}} \frac{g_{\chi}}{g_{* S}} x^{3 / 2} e^{-x}, 
\\
Y_{\cal B}^{\rm eq} & \simeq \frac{90}{(2 \pi)^{7 / 2}} \frac{g_{\cal B}}{g_{* S}}(2 x)^{3 / 2} e^{-2 x}e^{\left|{\cal E}_{\cal B}\right| / T},
\end{align}
where $g_\chi = 2$ and $g_{\cal B}$ are the dof of the $\chi$ particles and the bound states, with $g_{n\ell,s=0} = 2\ell+1$, $g_{n\ell,s=1} = 3(2\ell+1)$ for the spin singlet and triplet bound states respectively. Note that we lump together states of fixed $n,\ell,s$, but different $m_\ell$ and $m_s$. The annihilation and BSF cross-sections along with the rates $\Gamma_{\cal B}^{\rm dec},\Gamma_{\cal B}^{\rm ion}$ and $\Gamma_{{\cal B} \to {\cal B}'}^{\rm trans}$ which denote respectively the decay and ionisation rate of a bound level ${\cal B}$, and the transition of from the level ${\cal B}$ to $\mathcal{B^\prime}$, are defined in \cref{sec:FreezeOut_CrossSectionRates}. The angular brackets around cross-sections and bars above rates denote the thermally averaged quantities.  

For simplicity, we shall assume that the dark-photon temperature is the same as that of the Standard Model plasma, and include the dark-photon degrees of freedom in $g_*$ and $g_{*S}$. The dark-to-ordinary temperature ratio decreases at later times due to the decoupling of the Standard Model degrees of freedom, thereby ensuring that the extra radiation due to the dark photons satisfies existing constraints. If needed, a lower initial temperature for the dark plasma can be assumed with no impact on our conclusions.

\subsection{Cross-section and rates \label{sec:FreezeOut_CrossSectionRates}}

\subsubsection{Annihilation \label{sec:FreezeOut_CrossSectionRates_Ann}} 

\begin{figure}[t]
\centering
\begin{tikzpicture}[line width=1.1pt, scale=1]
\begin{scope}
\node at (-1,-2) {(a)};
\node at (-1.7, 1.3) {$\chi$};
\node at (-1.7,-0.3) {$\bar{\chi}$};
\node at (-2.2,0.45) {$\gamma$};
\draw[fermion] (-2,1) -- ( 0,1);
\draw[fermion] ( 0,1) -- ( 0,0);
\draw[fermion] ( 0,0) -- (-2,0);
\draw[vector] (-1.8,0) -- (-1.8,1);
\draw[vector] (-1.4,0) -- (-1.4,1);
\node at (-1,0.5) {$\cdots$};
\draw[vector] (-0.6,0) -- (-0.6,1);
\draw[vector] (0,1) -- (1,1);
\draw[vector] (0,0) -- (1,0);
\node at (1.3,1) {$\gamma $};
\node at (1.3,0) {$\gamma $};
\end{scope}
%
%
%
\begin{scope}[shift={(7,0)}]
\node at (-2.5,-2) {(b)};
\begin{scope}[shift={(-2.5,1.0)}]
\node at (-1.7, 1.3) {$\chi$};
\node at (-1.7,-0.3) {$\bar{\chi}$};
\node at (-2.2,0.45) {$\gamma$};
\node at (0.0,-0.5)  {$+$};
\draw[fermion] (-2,1) -- ( 0,1);
\draw[fermion] ( 0,0) -- (-2,0);
\draw[vector] (-1.8,0) -- (-1.8,1);
\draw[vector] (-1.4,0) -- (-1.4,1);
\node at (-1,0.5) {$\cdots$};
\draw[vector] (-0.6,0) -- (-0.6,1);
\draw[vector] (0,1) -- (0.4,1.7);
\node at (-0.2,1.4) {$\gamma $};
\draw[fermion] (0,1) -- (2,1);
\draw[fermion] (2,0) -- (0,0);
\draw[vector] (1.8,0) -- (1.8,1);
\draw[vector] (1.4,0) -- (1.4,1);
\node at      (1,0.5) {$\cdots$};
\draw[vector] (0.6,0) -- (0.6,1);
\draw[thin] (1.2,0.5) ellipse (0.95 and 0.85);
\end{scope}

\begin{scope}[shift={(-2.5,-1)}]
\node at (-1.7, 1.2) {$\chi$};
\node at (-1.7,-0.3) {$\bar{\chi}$};
\node at (-2.2,0.45) {$\gamma$};
\draw[fermion] (-2,1) -- ( 0,1);
\draw[fermion] ( 0,0) -- (-2,0);
\draw[vector] (-1.8,0) -- (-1.8,1);
\draw[vector] (-1.4,0) -- (-1.4,1);
\node at (-1,0.5) {$\cdots$};
\draw[vector] (-0.6,0) -- (-0.6,1);
\draw[vector] (0,0) -- (0.45,-0.7);
\node at (+0.0,-0.5) {$\gamma $};
\draw[fermion] (0,1) -- (2,1);
\draw[fermion] (2,0) -- (0,0);
\draw[vector] (1.8,0) -- (1.8,1);
\draw[vector] (1.4,0) -- (1.4,1);
\node at      (1,0.5) {$\cdots$};
\draw[vector] (0.6,0) -- (0.6,1);
\draw[thin] (1.2,0.5) ellipse (0.95 and 0.85);
\end{scope}
\end{scope}

\begin{scope}[shift={(7,0)}]
\node at (1.8,-2) {(c)};
\begin{scope}[shift={(2.5,1)}]
\draw[fermion] (-2,1) -- ( 0,1);
\draw[fermion] ( 0,1) -- ( 0,0);
\draw[fermion] ( 0,0) -- (-2,0);
\draw[vector] (-1.8,0) -- (-1.8,1);
\draw[vector] (-1.4,0) -- (-1.4,1);
\node at      (-1,0.5) {$\cdots$};
\draw[vector] (-0.6,0) -- (-0.6,1);
\draw[thin] (-1.2,0.5) ellipse (0.95 and 0.85);
\draw[vector] (0,1) -- (1,1);
\draw[vector] (0,0) -- (1,0);
\node at (1.3,1) {$\gamma $};
\node at (1.3,0) {$\gamma $};
\end{scope}
%
\begin{scope}[shift={(2.5,-1)}]
\draw[fermion] (-2,1)   -- ( 0,1);
\draw[fermion] ( 0,1)   -- ( 0,0.5);
\draw[fermion] ( 0,0.5) -- ( 0,0);
\draw[fermion] ( 0,0)   -- (-2,0);
\draw[vector] (-1.8,0) -- (-1.8,1);
\draw[vector] (-1.4,0) -- (-1.4,1);
\node at      (-1,0.5) {$\cdots$};
\draw[vector] (-0.6,0) -- (-0.6,1);
\draw[thin] (-1.2,0.5) ellipse (0.95 and 0.85);
\draw[vector] (0,1)   -- (1,1);
\draw[vector] (0,0.5) -- (1,0.5);
\draw[vector] (0,0)   -- (1,0);
\node at (1.3,1)   {$\gamma $};
\node at (1.3,0.5) {$\gamma $};
\node at (1.3,0)   {$\gamma $};
\end{scope}
\end{scope}

\end{tikzpicture}
\caption{
Feynman diagrams for (a) DM annihilation directly into dark photons, (b) DM capture into dark-positronium bound states, and (c) decay of dark-positronium into radiation.}
\label{fig:FeynmanDiagrams_AnnBSF}
\end{figure}
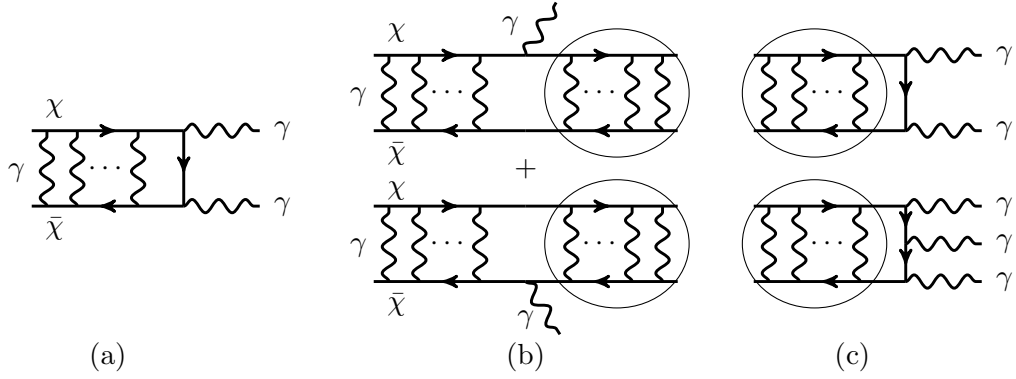

The dominant direct annihilation mode of $\chi\bar{\chi}$ pairs is into two dark photons, shown in \cref{fig:FeynmanDiagrams_AnnBSF},
\begin{align}
\chi + \bar{\chi}  \to \gamma + \gamma .
\label{eq:Ann}
\end{align}
In the non-relativistic regime, this is dominantly an $s$-wave process with tree-level cross-section times relative velocity
\begin{equation}
\sigma^{\rm ann}_{\rm tree} v_{\rm rel} = \frac{\pi \alpha^2}{m_\chi^2}  \equiv   \sigma_0.
\end{equation}
The attractive long-range  $\chi\bar{\chi}$ interaction due to dark-photon exchanges enhances the cross-section by the $s$-wave Sommerfeld factor $S^{\rm ann}(\zeta)$, with $\zeta \equiv \alpha / v_{\text{rel}}$, 
\begin{equation}
\label{eq:SEann-cross-section}
\sigma^{\text{ann}} v_{\text{rel}} = \sigma_0 S^{\rm ann}(\zeta),
\end{equation}
where \cite{Cassel:2009wt}
\begin{equation} 
\label{eq:SEfactor}
S^{\rm ann}(\zeta) = \frac{2 \pi \zeta}{1 - e^{-2 \pi \zeta}}.
\end{equation}
Assuming a Maxwellian distribution for the dark particles, the thermally averaged annihilation cross-section is
\begin{equation}
\left\langle\sigma^{\rm ann} v_{\rm rel}\right\rangle
= \sigma_0 \bar{S}^{\rm ann},
\end{equation}
where\footnote{\label{foot:z}
We note that in this simple model, $\bar{S}^{\rm ann}$ (cf.~\cref{eq:SEannthermallyavegared}), $\bar{S}_{\cal B}^{\rm BSF}$ (cf.~\cref{eq:SBSF_averaged}) and $f_{\cal B}^{\rm ion}$ (cf.~\cref{eq:fion}) can be expressed as functions of solely $z \equiv |\mathcal{E}_{\text{$n$=1}}|/T =\alpha^2 x/4$~\cite{vonHarling:2014kha}.}
\begin{align}
\label{eq:SEannthermallyavegared}
\bar{S}^{\rm ann}= & 
\frac{x^{3 / 2}}{2 \sqrt{\pi}} 
\int_0^{\infty} d v_{\rm rel} \,  v_{\rm rel}^2 
\, e^{-x v_{\rm rel}^2 / 4}
\, S^{\rm ann}_{\cal B}\left(\alpha / v_{\rm rel}\right).
\end{align}
$S^{\rm ann}$ and $\bar{S}^{\rm ann}$ are shown in \cref{fig:Sfactors}.

\subsubsection{Bound-state formation and ionisation}

Dark positronium forms with emission of a dark photon, 
\begin{align}
\chi + \bar{\chi} \to {\cal B} (\chi\bar{\chi}) + \gamma .
\label{eq:BSF}
\end{align}
The Feynman diagrams are shown in \cref{fig:FeynmanDiagrams_AnnBSF}.
At leading order in $\alpha$ and $v_{\rm rel}$, the BSF cross-sections do not depend on the spin configuration; the soft dark photon exchanges and ultrasoft dark photon emission imply that $\chi$ and $\bar{\chi}$ retain their spins individually. The velocity-weighted BSF cross-sections can be expressed as
\begin{equation}
\sigma^{\rm BSF}_{\cal B} v_{\rm rel}=\sigma_0 S^{\rm BSF}_{\cal B}(\zeta),
\end{equation}
where, averaging over $m_\ell$, the factors $S_{\cal B}^{\rm BSF}(\zeta)$ depend on the quantum numbers $n,\ell$; for the ground and first excited states~\cite{vonHarling:2014kha,Petraki:2015hla,Petraki:2016cnz,Harz:2018csl} 
\begin{subequations}
\begin{align}
S^{\rm BSF}_{10}   (\zeta) & = 
\frac{2^9}{3}    \frac{ \zeta ^4 e^{-4 \zeta  \cot ^{-1}(\zeta )}}{\left(\zeta ^2+1\right)^2} S^{\rm ann}(\zeta) ,
\\
S^{\rm BSF}_{20}  (\zeta ) & = 
\frac{2^{12}}{3} \frac{ \zeta^4\left(\zeta^2+1\right) e^{-4 \zeta  \cot ^{-1}\left(\frac{\zeta }{2}\right)}}{\left(\zeta ^2+4\right)^3} S^{\rm ann}(\zeta) ,
\\
S^{\rm BSF}_{21}  (\zeta ) & = 
\frac{2^{10}}{3} \frac{\zeta ^6 \left(11 \zeta ^2+12\right) e^{-4 \zeta  \cot ^{-1}\left(\frac{\zeta }{2}\right)}}{\left(\zeta ^2+4\right)^4} S^{\rm ann}(\zeta).
\end{align}
\end{subequations}
We iterate that $S_{\cal B}^{\rm BSF}$ are the same for any spin configuration, thus also for any averaging over them. 

The thermally-averaged velocity-weighted BSF cross-sections are
\begin{equation}
\left\langle\sigma^{\rm BSF}_{\cal B}v_{\rm rel}\right\rangle=\sigma_0 \bar{S}^{\rm BSF}_{\cal B}(z) ,
\end{equation}
where, assuming a Maxwellian velocity distribution for the dark particles,${}^{\ref{foot:z}}$
\begin{equation}
\bar{S}^{\rm BSF}_{\cal B} (z) = 
\frac{x^{3 / 2}}{2 \sqrt{\pi}} 
\int_0^{\infty} d v_{\rm rel} 
\, v_{\rm rel}^2 
\, e^{-x v_{\rm rel}^2 / 4}
\, S^{\rm BSF}_{\cal B}\left(\alpha / v_{\rm rel}\right) 
\left[1+f_\gamma(\omega)\right].
\label{eq:SBSF_averaged}
\end{equation}
The factor $1 + f_\gamma(\omega)$ above is the Bose enhancement due to the final-state dark photon, with $f_\gamma(\omega)$ being the dark photon distribution and $\omega$ being the dissipated energy~\cite{Petraki:2015hla} 
\begin{align}
\omega \simeq \dfrac{1}{2}\mu v_{\rm rel}^2 + |{\cal E}_{\cal B}|. 
\label{eq:omega}
\end{align} 
For the thermal dark photons, $f_\gamma(\omega)=1/(e^{\omega / T}-1)$, and the Bose factor is significantly larger than 1 for $\omega \lesssim T$. Upon thermally averaging, this corresponds to $T \gtrsim |{\cal E}_{\cal B}|$; in this regime, the Bose factor must be included in order to ensure detailed balance between BSF and ionisation processes~\cite{vonHarling:2014kha}.

\begin{figure}[t!]
\centering
\includegraphics[width=0.98\linewidth] {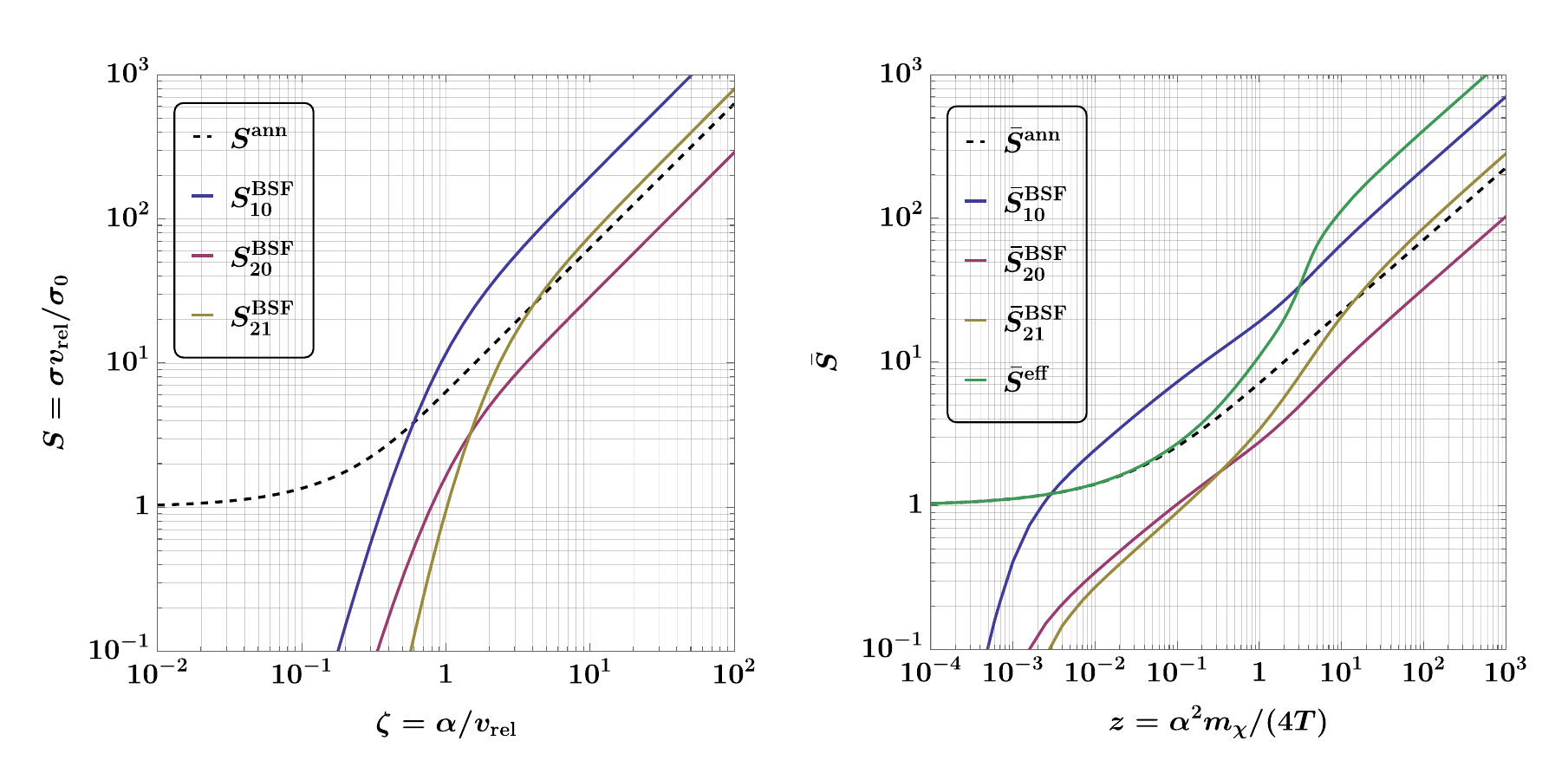}
\caption{(a) $S^{\rm ann}$ and $S^{\rm BSF}_{\cal B}$ as a function of $\zeta=\alpha / v_{\rm rel}$. (b) The thermally averaged $\bar{S}^{\rm ann}$ and $\bar{S}^{\rm BSF}$ for the ground and first excited states as a function of $z= m_\chi \alpha^2 /4 T$. The effective factor, $\bar{S}^{\rm eff}$ is plotted for a dark particle of mass $m_\chi=10~{\rm TeV}$ with the corresponding required coupling to obtain the observed DM abundance by taking into account both SE annihilations and BSF.}
\label{fig:Sfactors}
\end{figure}

The BSF factors $S_{\cal B}^{\rm BSF}$ and $\bar{S}_{\cal B}^{\rm BSF}$ are shown in \cref{fig:Sfactors}, and compared with the corresponding annihilation factors. While BSF is suppressed at large velocities, it becomes faster than direct annihilation at $v_{\rm rel} \lesssim \alpha$. Both annihilation and BSF scale as $S\propto \zeta =\alpha/v_{\rm rel}$ in this regime, owing to the Sommerfeld enhancement due to the attractive long-range interaction between the incoming particles. Capture into the ground state is the dominant BSF process, while for the first excited level, $n=2$, capture into the $\ell=1$ states dominates over capture into the $\ell=0$ state, in large part due to the higher multiplicity of the former.

While the comparison between $S^{\rm ann}$ and $S_{\cal B}^{\rm BSF}$ suggests that upon thermal averaging, BSF should dominate over annihilation at $T \lesssim |{\cal E}_{\cal B}|$, this in fact occurs already at much higher temperatures due to the Bose enhancement factor. Nevertheless, it does not transcribe into faster DM depletion via BSF rather than direct annihilation throughout, due to the bound-state ionisation processes, as we discuss in the following.

\subsubsection{Ionisation rates}

Bound states can get dissociated by the inverse of process \eqref{eq:BSF}. The ionisation cross-section is related to the BSF one via the Milne relation (see e.g.~\cite[appendix~D]{Harz:2018csl} and \cref{App:CrossSections}),
\begin{align}
\sigma_{\cal B}^{\rm ion}
=\frac{g_{\chi}^2}{g_{\gamma}g_{\cal B}} 
\frac{\mu^2 v_{\rm rel}^2}{\omega^2}
\, \sigma_{\cal B}^{\mathsmaller{\rm BSF}}.
\label{eq:Milne}
\end{align}
The ionisation rate of a bound level ${\cal B}$ is obtained by averaging over the dark photon distribution
\begin{equation}
\bar{\Gamma}_{\cal B}^{\rm ion}
=g_\gamma \int \frac{d^3 p}{(2 \pi)^3} f_\gamma(\omega) \sigma_{\cal B}^{\text {ion }},
\label{eq:GammaIonB_def}
\end{equation}
where we recall that $\omega$ is given by \cref{eq:omega}.  
Considering the thermal distribution of dark photons, this becomes
\begin{equation}
(\bar{\Gamma}_{\cal B}^{\rm ion})_{\rm thermal} 
= \mu \alpha^5 \, f^{\text{ion}}_{\cal B},
\label{eq:GammaIonB_thermal}
\end{equation}
where${}^{\ref{foot:z}}$
\begin{equation}
f^{\text{ion}}_{\cal B} = 
\dfrac{1}{8 \pi} \int^{\infty}_{0} 
\dfrac{d \zeta \, \zeta^{-4}}
{\exp\left[\left(\dfrac{1}{n^2} + \dfrac{1}{\zeta^2}\right) \dfrac{x\alpha^2}{4} \right] -1} 
\ S^{\rm BSF}_{\cal B} (\zeta).
\label{eq:fion}
\end{equation}
Alternatively, $(\bar{\Gamma}_{\cal B}^{\rm ion})_{\rm thermal}$ can be calculated from the thermally averaged BSF rate by invoking detailed balance,  
\begin{equation}
(\bar{\Gamma}_{\cal B}^{\rm ion})_{\rm thermal} 
= s\left\langle\sigma_{ {\cal B}}^{\rm BSF} v_{\rm rel}\right\rangle 
\left[ (Y^{\rm eq}_{\chi})^2 / Y_{\cal B}^{\rm eq}\right]. 
\label{eq:GammaIon_thermal_DetailedBalance}
\end{equation}

\subsubsection{Bound-to-bound transitions}

Similarly to BSF and ionisation, bound-to-bound transitions occur via emission or absorption of dark photons. 
The selection rules at leading order are $\Delta \ell = \pm 1$, $\Delta s=0$~\cite{Petraki:2016cnz},
\begin{align}
{\cal B} \leftrightarrow {\cal B}' + \gamma .
\label{eq:B2B}
\end{align}
The only allowed transitions for the bound levels we are considering are thus between the $n=2, \ell=1$ and $n=1, \ell=0$ states of the same spin. The rate of de-excitation $\{n=2,\ell=1\} \to \{n=1,\ell=0\}+\gamma$, averaged over the $m_\ell$ values of the excited state, is (see e.g.~\cite{Biondini:2023zcz})
\begin{align}
\Gamma^{\rm trans}_{\mathsmaller{ \{21\} \rightarrow \{1 0\}} } 
= \frac{2^8}{3^8} \mu \alpha^5. 
\label{eq:Gamma_21to10}
\end{align}
In thermally averaging \cref{eq:Gamma_21to10}, we must take into account the Bose enhancement factor due to the final-state dark photon, emitted with energy equal to the difference between the binding energies of the two levels, $\mathcal{E}_{21} - \mathcal{E}_{10} = 3\mu \alpha^2/8$. On the other hand, since the ensemble of bound states are non-relativistic, we may ignore any time dilation factor. Thus
\begin{align}
\bar{\Gamma}^{\rm trans}_{\mathsmaller{ \{21\} \rightarrow \{1 0\}} } 
= \frac{2^8}{3^8} \mu \alpha^5 \times 
\left(1+\dfrac{1}{e^{3x \alpha^2/16} - 1}\right).
\label{eq:Gamma_21to10_average}
\end{align}
The inverse transition rate can be found from detailed balance, 
$\bar{\Gamma}_{{\cal B} \rightarrow {\cal B}^{\prime}}^{\rm trans}
=\bar{\Gamma}_{{\cal B}^{\prime} \rightarrow {\cal B}}^{\rm trans}
\left(Y_{{\cal B}^{\prime}}^{\rm eq} / Y_{\cal B}^{\rm eq}\right)$, 
which in the case at hand implies
\begin{align}
\bar{\Gamma}^{\rm trans}_{\mathsmaller{ \{10\} \rightarrow \{21\}} }     
= \frac{2^8}{3^8} \mu \alpha^5 \times 
\dfrac{1}{e^{3x \alpha^2/16} - 1}.
\label{eq:Gamma_10to21_average}
\end{align}

\subsubsection{Bound-state decay}

The bound states decay into an odd or even number of dark photons according to the selection rule  $(-1)^{\ell + s} = (-1)^{n_{\gamma}}$, arising from charge conjugation invariance, where $n_\gamma$ is the number of dark photons.
We shall denote the spin-singlet ($s=0$) states with $S$ and the spin-triplet ($s=1$) with $T$.

The states 
$\{n=1,\ell=0;~s=0\}$, 
$\{n=2,\ell=0;~s=0\}$ and 
$\{n=2,\ell=1;~s=1\}$ 
decay into two dark photons with rates~\cite{Stroscio:1975fa}
\begin{subequations}
\begin{align}
\Gamma^{\rm dec}_{10,S} &= \mu \alpha^5, \\
\Gamma^{\rm dec}_{20,S} &= \dfrac{\mu \alpha^5}{8}, \\
\Gamma^{\rm dec}_{21,T} &= \dfrac{\mu \alpha^7}{160}.
\end{align}
\end{subequations}
The states 
$\{n=1,\ell=0;~s=1\}$, 
$\{n=2,\ell=0;~s=1\}$ and 
$\{n=2,\ell=1;~s=0\}$ 
decay into three dark photons with rates~\cite{Stroscio:1975fa}
\begin{subequations}
\begin{align}
\Gamma^{\rm dec}_{10,T} 
&= \frac{4\left(\pi^2-9\right)}{9 \pi}  \mu \alpha^6,  
\\
\Gamma^{\rm dec}_{20,T}
&= \frac{\pi^2-9}{18 \pi} \mu \alpha^6, 
\\
\Gamma^{\rm dec}_{21,S}
&=\frac{\mu \alpha^8 \ln (32/\alpha^2) } {48 \pi}.
\end{align}
\end{subequations}
In the dynamical system we are considering, bound states are non-relativistic, thus time dilation is negligible and the decay rates listed above (evaluated at the bound-state rest frame) approximate well the thermally averaged ones.

\subsection{Effective Boltzmann equation \label{sec:FreezeOut_EffectiveBoltzmann}}

Considering the coupled equations of \cref{sec:FreezeOut_Boltzmann} and the rates summarised in \cref{sec:FreezeOut_CrossSectionRates}, the calculation of the DM freeze-out in this system requires in principle integrating a set of seven coupled Boltzmann equations for the free and bound species. However, metastable bound states typically attain a quasi-steady state in an expanding universe: at early times, BSF and ionisation are very efficient, while at later times bound-state decays into radiation and/or de-excitations to lower lying states exceed the expansion rate of the universe. This implies that $d Y_{\cal B}/dx \simeq 0$ at all relevant times, yielding algebraic equations for $Y_{\cal B}$ in terms of the densities of the free species and the various interaction rates involved. These equations generalise the Saha equilibrium to metastable bound states, and we further discuss them in \cref{sec:FreezeOut_GenSahaEquil}. They can be re-deployed to reduce the coupled system into a single effective equation~\cite{Ellis:2015vaa,Binder:2021vfo}
\begin{equation}
\label{eq:BoltzmannEffective}
\frac{d Y_\chi}{d x}=-\frac{\lambda}{x^2} 
\langle\sigma^{\rm eff} v_{\rm rel}\rangle
\left[Y_{\chi}^{2}-\left(Y_{\chi}^{\rm eq}\right)^2\right],
\end{equation} 
where we note that the attractor solution remains the $\chi$ equilibrium density~\cite{Binder:2021vfo}, and the effective thermally averaged cross-section is\footnote{If the assumption that at least one interaction rate is large at any given time fails for a bound state, then this also means that this state does not contribute significantly to the DM depletion due to small BSF and/or decay and transition rates. Including or removing it from the equations has no effect. It is therefore essentially always safe to make the quasi-steady state approximation for reducing the coupled system to a single differential equation, even though the generalised Saha equilibrium equation (c.f.~\cref{eq:SahaGen}) may not be sufficiently accurate for the density of that bound state~\cite{Binder:2021vfo}.} 
\begin{equation}
\label{eq:EffectiveCrossSection}
\langle\sigma^{\rm eff} v_{\rm rel}\rangle 
\equiv 
\langle \sigma^{\rm ann} v_{\rm rel}\rangle + 
\sum_{\cal B} r_{\cal B} 
\langle\sigma_{\cal B}^{\rm BSF} v_{\rm rel}\rangle.
\end{equation}
In \cref{eq:EffectiveCrossSection}, every BSF cross-section is weighted by an efficiency factor $r_{\cal B} \in [0,1]$ that encapsulates the interplay of ionisation, decay, and transitions processes. Organising these rates appropriately in matrices, the factors $r_{\cal B}$ can be expressed in closed form for an arbitrary number of bound levels~\cite{Binder:2021vfo}. For our model, they are
\begin{subequations}
\label{eq:EfficiencyFactors}
\label[pluralequation]{eqs:EfficiencyFactors}
\begin{align}
r_{10}  &= \frac
{\bar{\Gamma}^{\rm dec}_{21} \bar{\Gamma}^{\rm trans}_{ \{10\} \rightarrow \{21\}} + 
 \bar{\Gamma}^{\rm dec}_{10} \left (\bar{\Gamma}^{\rm dec}_{21} + \bar{\Gamma}^{\rm ion}_{21}+ \bar{\Gamma}^{\rm trans}_{ \{21\} \rightarrow \{10 \}}  \right) }
{
\left(\bar{\Gamma}^{\rm dec}_{21} + \bar{\Gamma}^{\rm ion}_{21} \right)  
\left(\bar{\Gamma}^{\rm dec}_{10} + \bar{\Gamma}^{\rm ion}_{10} + \bar{\Gamma}^{\rm trans}_{ \{10\} \rightarrow \{21\}}  \right)  
+ 
(\bar{\Gamma}^{\rm dec}_{10} + \bar{\Gamma}^{\rm ion}_{10} ) \bar{\Gamma}^{\rm trans}_{ \{21\} \rightarrow \{10 \}} }, 
\label{eq:EfficiencyFactor_r10}
\\
r_{20}  & = \frac{\bar{\Gamma}^{\rm dec}_{20}}{\bar{\Gamma}^{\rm dec}_{20} + \bar{\Gamma}^{\rm ion}_{20}}, 
\label{eq:EfficiencyFactor_r20}
\\
r_{21} & = \frac
{\bar{\Gamma}^{\rm dec}_{21} \left(\bar{\Gamma}^{\rm dec}_{10} + \bar{\Gamma}^{\rm ion}_{10} +  \bar{\Gamma}^{\rm trans}_{ \{10\} \rightarrow \{21 \}} \right)  
+\bar{\Gamma}^{\rm dec}_{10}   \bar{\Gamma}^{\rm trans}_{ \{21\} \rightarrow \{10\}} } 
{\left(\bar{\Gamma}^{\rm dec}_{21} + \bar{\Gamma}^{\rm ion}_{21} \right)  
 \left(\bar{\Gamma}^{\rm dec}_{10} + \bar{\Gamma}^{\rm ion}_{10} + \bar{\Gamma}^{\rm trans}_{\{10\}\rightarrow\{21 \}} \right)  
+ (\bar{\Gamma}^{\rm dec}_{10} + \bar{\Gamma}^{\rm ion}_{10} ) 
\bar{\Gamma}^{\rm trans}_{ \{21\} \rightarrow \{10 \}} }.
\label{eq:EfficiencyFactor_r21}
\end{align}
\end{subequations}
We emphasise that all the rates above are thermally averaged, and the factors $r_{\cal B}$ are different for the spin-singlet and spin-triplet states due to the different bound-state decay rates.

\subsection{Relic density \label{sec:FreezeOut_RelicDensity}}

\begin{figure}[t!]
\centering
\includegraphics[width=0.98\linewidth] {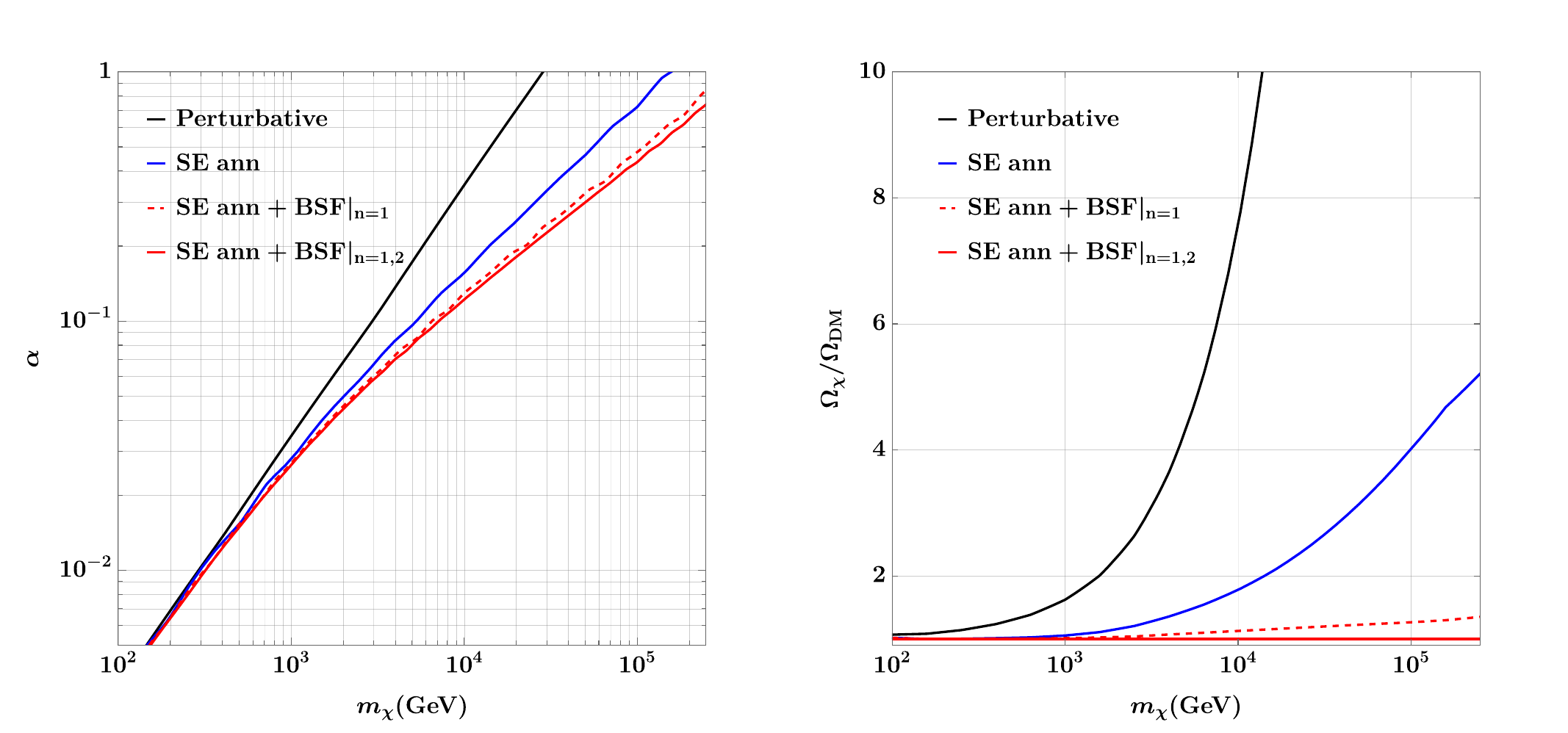}
\caption{\textit{Left}: The required coupling $\alpha$ to obtain the observed DM density as a function of the dark particle mass $m_{\chi}$. We consider perturbative annihilation only (black), SE annihilation (blue), SE annihilation + BSF of the ground state only (dotted red line), and of ground+first excited state (solid red line).
\textit{Right}: For the combination of $\left\{m_\chi, \alpha\right\}$ values predicted by the full calculation with the first excited state (solid red line on the left), we show the ratio of the relic density $\Omega_\chi$ to the observed DM density in different approximations. This indicates the magnitude of each effect on the relic density. 
}
\label{fig:FreezeOut_Results}
\end{figure}

Integrating the effective Boltzmann \cref{eq:BoltzmannEffective}, we find the final value of the comoving density $Y_\chi(\infty)$. The fractional DM relic density is 
\begin{equation}
\Omega_\chi = \frac{2 m_{\chi} Y_{\chi}(\infty) s_{0}}{\rho_{c}},
\end{equation}
where the factor of 2 accounts for both $\chi$ and $\bar{\chi}$ and $s_0 \simeq 2839.5$ cm$^{-3}$, and $\rho_c = 4.78 10^{-6}$ GeV cm$^{-3}$ are the present-day entropy and critical energy density of the Universe \cite{Aghanim:2018eyx}. 
Requiring that  $\Omega_{\chi} = \Omega_{\rm DM} \simeq 0.26$, we determine the relation between $\alpha$ and $m_\chi$ at different approximations: considering direct DM annihilation only at tree level, including the Sommerfeld enhancement (SE) of the direct annihilation, incorporating capture into the ground state and including also the first excited state. Assuming then the mass-coupling relation from the last and most accurate calculation, we estimate the effect of each process on the DM relic density. Our results are shown in \cref{fig:FreezeOut_Results}. BSF decreases the DM density by up to a factor of about 5, with the first excited state accounting for a factor of about 1.2 at $m_\chi \gtrsim 100$~TeV. We use the mass-coupling relation from the most complete calculation in the computations that follow. 

We note that in the present model, the DM depletion is dominated by the $s$-wave annihilation into two dark photons and $p$-wave capture into the ground state. Higher partial waves, contributing mostly via capture into excited states, are subdominant. Therefore, the relevant mass range goes up to the combined $s+p$ unitarity limit, i.e.~$m_\chi \lesssim 280$~TeV~\cite{Baldes:2017gzw}.

\subsection{Generalised Saha equilibrium \label{sec:FreezeOut_GenSahaEquil}}

\begin{figure}[t!]
\centering
\includegraphics[width=0.98\textwidth] {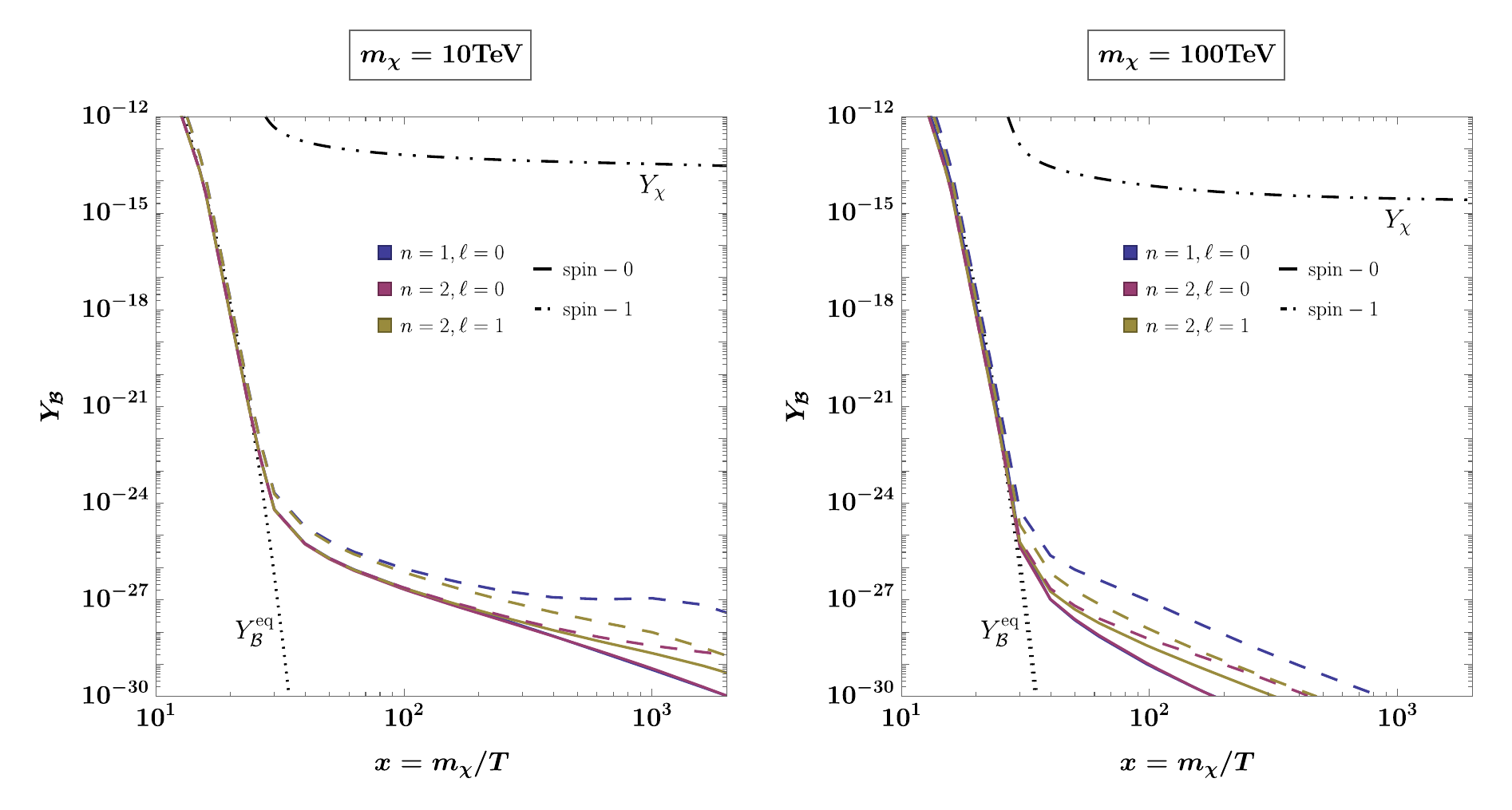}
\caption{Bound-state comoving densities, $Y_{\cal B}$, computed from the generalised Saha equilibrium \cref{eq:SahaGen} for the ground and excited states, for two indicative DM masses $m_{\chi} = 10$ and $100$ TeV with the required coupling to reproduce observed DM density. The solid (dashed) lines correspond to the spin singlet (triplet) bound states. The bound-state equilibrium density for the ground state singlet and the density of the free particles are also plotted. Although the bound-state densities deviate from the equilibrium ones after DM freeze-out, they remain orders of magnitude smaller than the density of the free species due to the bound-state decay. Nevertheless, despite the decay, the bound state densities do not return to their equilibrium values at late times, having inherited in part the chemical potential of the free species. (See~\cref{sec:FreezeOut_GenSahaEquil} for discussion.)
}
\label{fig:bsdensity}
\end{figure}

The quasi-steady-state approximation discussed in \cref{sec:FreezeOut_EffectiveBoltzmann} yields algebraic equations that relate the densities of the bound states to those of the free species via the various interaction rates involved. These equations generalise the well-known Saha equilibrium that pertains to stable bound states, to metastable bound states, they offer insight into the relevant dynamics, and will be essential in computing the back-reaction of radiation produced during the DM thermal decoupling in the next section. For these reasons we discuss them here in conjunction with the freeze-out dynamics described in the previous sections. 

The generalised Saha equilibrium reads~\cite{Binder:2021vfo}
\begin{equation}
\frac{Y_{\cal B}}{Y_{\cal B}^{\rm eq}} =
(1-r_{\cal B}) \frac{Y_{\chi}^2}{(Y_{\chi}^{\rm eq})^2} + r_{\cal B},
\label{eq:SahaGen}
\end{equation}
where the efficiency factors $r_{\cal B} \in [0,1]$ can be expressed in closed form in terms of the bound-to-bound transition, bound-state ionisation and decay rates, appropriately organised in matrices, even for an arbitrary number of bound states.\footnote{
More generally, if the bound states consist of $N$ particles $\chi_1 \cdots \chi_N$, the factor  $(Y_{\chi} / Y_{\chi}^{\rm eq})^2$ in \cref{eq:SahaGen}  should be replaced by 
$(Y_{\chi_1} \cdots Y_{\chi_N}) / (Y_{\chi_1}^{\rm eq} \cdots Y_{\chi_N}^{\rm eq})$.} 
We refer to \cite{Binder:2021vfo} for the exact formulae, and quote here the result for the simple case when there are no bound-to-bound transitions,
\begin{align}
r_{\cal B} = \dfrac{\bar{\Gamma}_{\cal B}^{\rm dec}}
{\bar{\Gamma}_{\cal B}^{\rm dec} + \bar{\Gamma}_{\cal B}^{\rm ion}} .
\label{eq:rB_NoTrans}
\end{align}
In the opposite limit of very rapid bound-to-bound transitions, all $r_{\cal B}$ factors become equal and are given by \cref{eq:rB_NoTrans} with the decay and ionisation rates replaced by the corresponding rates averaged over all bound states~\cite{Binder:2021vfo}. Quite generally the factors $r_{\cal B}$ vanish if the bound-state decay rates vanish.

We may discern two formal limits in \cref{eq:SahaGen}: 
(i) if $r_{\cal B} = 0$, then 
$Y_{\cal B} / Y_{\cal B}^{\rm eq} = (Y_{\chi} / Y_{\chi}^{\rm eq})^2$, 
which reproduces the standard Saha equilibrium for stable bound states, and 
(ii) if $r_{\cal B} = 1$, then $Y_{\cal B} / Y_{\cal B}^{\rm eq} = 1$, i.e.~the bound states attain their equilibrium distribution, as expected for species whose decay rate exceeds the expansion rate of the universe.

Let us now explore different regimes in the context of the DM thermal freeze-out, keeping in mind the above. 
An illustration of the bound-state comoving densities is presented in \cref{fig:bsdensity}. 
\begin{itemize}
\item 
At temperatures much higher than the binding energies, $T \gg |{\cal E}_{\cal B}|$, the bound-state ionisation rates are typically much larger than the decay rates, $\Gamma_{\cal B}^{\rm ion} \gg \Gamma_{\cal B}^{\rm dec}$, setting $r_{\cal B} \ll 1$ (cf.~\cref{eq:rB_NoTrans}) and realising the limit (i) above. 

In this regime, the free species are in equilibrium before freeze-out, the same thus follows for the bound states. After freeze-out, the free species develop a chemical potential that is inherited by the bound states, whose densities consequently start becoming much larger than their equilibrium values.  

\item
As the temperature falls around or below the binding energies, $T \lesssim |{\cal E}_{\cal B}|$, the ionisation rates decrease exponentially fast, eventually falling below the bound-state decay rates and driving $r_{\cal B} \to 1$. Focusing on the no-transition case for simplicity, 
$1-r_{\cal B} = \bar{\Gamma}_{\cal B}^{\rm ion} / \bar{\Gamma}_{\cal B}^{\rm dec}$; then, 
using the detailed balance \cref{eq:GammaIon_thermal_DetailedBalance}, the generalised Saha equilibrium \cref{eq:SahaGen} yields
\begin{align}
Y_{\cal B} \simeq  \left(
\dfrac{\langle \sigma_{\cal B}^{\rm BSF} v_{\rm rel} \rangle s }{\bar{\Gamma}_{\cal B}^{\rm dec}} \right) Y_{\chi}^2 
+ Y_{\cal B}^{\rm eq} .
\label{eq:SahaGen_Late}
\end{align}
\Cref{eq:SahaGen_Late} remains valid in the presence of rapid bound-to-bound transitions if we replace 
$\langle \sigma_{\cal B}^{\rm BSF} v_{\rm rel} \rangle$ and $\bar{\Gamma}_{\cal B}^{\rm dec}$ with the corresponding values averaged over all bound levels. 

Independently of the detailed form of $r_{\cal B}$, the first term in \cref{eq:SahaGen_Late} continues to dominate over the second, owing to the large chemical potential of $Y_{\chi}$; consequently, $Y_{\cal B}$ remains significantly larger than $Y_{\cal B}^{\rm eq}$ despite $r_{\cal B} \to 1$. In the present model, where the attractive long-range interaction in the scattering state sets $\langle\sigma_{\cal B}^{\rm BSF} v_{\rm rel} \rangle \propto T^{-1/2}$ at $T\lesssim|{\cal E}_{\cal B}|$, \cref{eq:SahaGen_Late} implies that the bound-state densities decrease approximately as $Y_{\cal B} \propto T^{5/2}$ at low temperatures. Even in models with Sommerfeld suppressed BSF cross-sections due to a repulsive interaction in the scattering state, 
$\langle \sigma_{\cal B}^{\rm BSF} v_{\rm rel} \rangle \propto (|{\cal E}_{\cal B}|/T)^{1/2} \exp (-2\pi\sqrt{|{\cal E}_{\cal B}|/T})$ at $T\lesssim|{\cal E}_{\cal B}|$, 
thus the first term in \cref{eq:SahaGen_Late} still does not decrease at low temperatures as quickly as $Y_{\cal B}^{\rm eq}$, and never attains this value.

Nevertheless, $Y_{\cal B}$ remains much smaller than $Y_{\chi}$, even though BSF may be very fast and / or the entropy of the plasma may be sufficiently low to render the capture into bound states energetically favourable. This is in sharp contrast to the case of stable bound states, whose density increases exponentially when the temperature drops, at least as long as the capture processes remain fast enough to overcome the expansion of the universe (cf.~Hydrogen recombination). This distinction is crucial in estimating the back-reaction of radiation produced during the DM thermal decoupling in the next section.

All in all, we find $Y_{\cal B}^{\rm eq} \ll Y_{\cal B}  \ll  Y_{\chi}$ at low $T$.   
\end{itemize}

\section{The fate of dark radiation produced during DM thermal decoupling \label{Sec:PhotonDist}}

To fully assess whether the radiation produced during the DM thermal decoupling back-reacts, ionising or exciting bound states and hindering the DM depletion via BSF, requires solving a set of coupled equations for the phase-space density of the dark radiation, the free and the bound DM species. This is computationally quite demanding. We instead follow another approach. 

We begin by formulating the Boltzmann equation for the phase-space density of the dark photons. Using the results of the previous section for the densities of the free and bound DM particles (which assume kinetic equilibrium and ignore back-reaction by the radiation produced during the DM thermal decoupling), we estimate the rates of the various processes to which the dark photons produced during the DM thermal decoupling may participate. 
We show that both the resonant and the higher-energy dark photons are far more likely to thermalise or redshift, than ionise or excite the metastable bound states. 
This demonstrates unambiguously that the backreaction of the produced radiation is negligible.

\subsection{Boltzmann equation for the dark photon distribution  \label{sec:PhotonDist_BoltzmannEq}}

The evolution of the dark photon energy distribution, $f_{\gamma} (\omega,t)$, is governed by the equation $L = C$, where $L$ is the Liouville operator and $C$ encompasses elastic and inelastic collision terms. The Liouville operator is given by
\begin{equation}
\label{eq:LiouvilleOp_def}
L = \omega \left[\frac{\partial f_{\gamma} (\omega,t)}{\partial t} 
- H \omega \frac{\partial f_{\gamma} (\omega,t)}{\partial \omega} \right] 
\end{equation}
where $H$ is the Hubble parameter. Defining the dimensionless comoving energy
\begin{align}
\varpi \equiv \omega / T,
\label{eq:ConformalVariables}
\end{align}
and using the time variable $x=m_{\chi}/T$, as for freeze-out, the Liouville operator for the dark photon phase space density $f_{\gamma}(\omega,t) = F_{\gamma}(\varpi,x)$ becomes 
\begin{equation}
    L =  m_{\chi} \varpi 
\, \frac{1}{\Lambda x^2}
\, \frac{\partial F_{\gamma}(\varpi,x)}{\partial x} ,
\label{eqs:LiouvilleOp_final}
\end{equation}
where $\Lambda$ is defined in \cref{eq:Lambda}. 
The elastic and inelastic collision terms are
\begin{subequations}
\label{eq:CollisionTerms}
\label[pluralequation]{eqs:CollisionTerms}
\begin{align}
C = C^{\rm elas} 
+ \sum_{\cal B} C_{\cal B}^{\rm BSF}
+ \sum_{{\cal B}\neq {\cal B}'} \sum_{{\cal B}'} C_{{\cal B}{\cal B}'}^{\rm B2B}
+ \sum_{\cal B} C_{\cal B}^{\rm BSD} 
+ C^{\rm ann} ,
\tag{\ref{eqs:CollisionTerms}}
\label{eq:CollisionOp}
\end{align}
with
\begin{align}
C^{\rm elas} & =- \frac{2}{2g_{\gamma}} \sum_{{\rm dof}} 
\int d\Pi_{\chi}  d\Pi_{\chi}'  d\Pi_{\gamma}' 
\left\{
f_{\gamma}(\omega,t)f_{\chi}(E_{\chi}) [1+f_{\gamma} (\omega',t)] - f_{\gamma}(\omega',t)f_{\chi}(E_{\chi}') [1+f_{\gamma} (\omega,t)]
\right\} \nonumber \\ 
& \times 
|\mathcal{M}^{\rm elas}|^2 \, (2\pi)^4 \delta^4(p_{\gamma} + p_{\chi} - p_{\gamma}' - p_{\chi}') ,
\label{eq:Collision_elas} 
\\
C_{\cal B}^{\rm BSF} &=- \frac{1}{2g_{\gamma}} \sum_{{\rm dof}} 
\int d\Pi_{\chi} \, d\Pi_{\bar{\chi}} \, d\Pi_{\cal B} 
\left\{
f_{\gamma}(\omega,t) \, f_{\cal B} (E_{\cal B}) - f_{\chi}(E_{\chi})\, f_{\chi}(E_{\bar{\chi}}) [1+f_{\gamma}(\omega,t)]
\right\} \nonumber \\ 
&\times 
|\mathcal{M}_{\cal B}^{\rm BSF}|^2 \, (2\pi)^4 \delta^4(p_{\gamma} + p_{\cal B} - p_{\chi} - p_{\bar{\chi}}) ,
\label{eq:Collision_BSF}
\\
C_{{\cal B}{\cal B}'}^{\rm B2B} &= - \frac{1}{2g_{\gamma}} \sum_{{\rm dof}} 
\int d\Pi_{\cal B} \, d\Pi_{{\cal B}'} 
\left\{
f_{\cal B}(E_{\cal B}) f_{\gamma}(\omega,t) -f_{{\cal B}'} (E_{\mathcal{B'}}) \, [1+f_{\gamma} (\omega,t)]
\right\} \nonumber \\ 
& \times 
|\mathcal{M}_{{\cal B}{\cal B}'}^{\rm B2B}|^2 \, (2\pi)^4 \delta^4(p_{\gamma}+p_{\cal B} - p_{{\cal B}'}) ,
\label{eq:Collision_B2B}
\\
C_{\cal B}^{\rm BSD} &= - \frac{1}{2g_{\gamma}} \sum_{\rm dof} 
\int d\Pi_{\cal B} \, d\Pi_{\gamma}'
\left\{ f_{\gamma}(\omega,t) f_{\gamma}(\omega',t) - f_{\cal B}(E_{\cal B}) \right\} \nonumber \\
& \times
|\mathcal{M}^{\rm BSD}|^2 \, (2\pi)^4 \delta^4(p_{\gamma}+p_{\gamma}'-p_{\cal B}),
\label{eq:Collision_BSD2}
\\
C^{\rm ann} &= - \frac{1}{2g_{\gamma}} \sum_{{\rm dof}} 
\int d\Pi_{\chi} \, d\Pi_{\bar{\chi}} \, d\Pi_{\gamma}'
\left\{ f_{\gamma}(\omega,t) f_{\gamma}(\omega',t) - f_{\chi}(E_{\chi}) f_{\chi}(E_{\bar{\chi}}) \right\} \nonumber \\ 
& \times
|\mathcal{M}^{\rm ann}|^2 \, (2\pi)^4 \delta^4(p_{\gamma}+p_{\gamma}'-p_{\chi}-p_{\bar{\chi}}),
\label{eq:Collision_Ann}
\end{align}
\end{subequations}
where $d\Pi_j = \frac{d^3p_j}{(2\pi)^3 2E_j} $ and the extra factor 2 in \eqref{eq:Collision_elas} with respect to all other collision terms accounts for scattering of dark photons on both $\chi$ and $\bar{\chi}$. No additional factor is required in \cref{eq:Collision_Ann,eq:Collision_BSD2} where there are two identical particles in the initial state~\cite{Gondolo:1990dk}. The sums denote summation over the dof of all initial and final state particles. We calculate the expressions for the collision terms next.

\subsection{Collision terms \label{sec:PhotonDist_CollisionTerms}}

In this section, we carry out the phase-space integrations in the collision terms \eqref{eqs:CollisionTerms}, to bring them in a convenient form. We will make use of some standard formulae collected in \cref{App:CrossSections}.

\subsubsection{Elastic scattering \label{sec:PhotonDist_CollisionTerms_ElasticScatt}}

The dark photons thermalise mostly via scattering elastically on free DM particles,
\begin{align}
\chi + \gamma \leftrightarrow  \chi + \gamma   .
\end{align}
For low-energy dark photons $\omega \ll m_\chi$, the collision term \eqref{eq:Collision_elas} can be expanded in powers of the dark photon energy transfer yielding a Fokker-Planck-type term. For the case at hand, this is the Kompaneets equation~\cite{Kompaneets:1957,Challinor_1998x}
\begin{align}
C^{\rm elas} 
&\simeq 2
m_{\chi} \, n_{\chi} \sigma_{T} \, \frac{1}{x^2 \, \varpi} 
\frac{d}{d\varpi} \left[\varpi^4  j(\varpi, x) \right],
\label{eq:Celas_KompaneetsEq}
\end{align}
where $\sigma_T = \frac{8\pi\alpha^2}{3 m_{\chi}^2}$ is the Thomson cross-section, the factor 2 accounts for dark photon scattering on both $\chi$ and $\bar{\chi}$, and the current $j(\varpi, x )$ is
\begin{align}
\label{Kompaneetscurrent}
j(\varpi, x )  
&=   \left(\frac{dF_{\gamma}}{d\varpi} + F_{\gamma} + F_{\gamma}^2\right) 
\\
&+ \frac{1}{x} \Bigg[ 
\frac{5}{2}  \left(\frac{dF_{\gamma}}{d\varpi} + F_{\gamma} + F_{\gamma}^2\right) 
+ \frac{21}{5} \varpi 
\left(\frac{d^2 F_{\gamma}}{d\varpi^2} + \frac{dF_{\gamma}}{d\varpi} (2F_{\gamma}+1) \right) 
\nonumber \\
&+ \frac{7}{10} \varpi^2 \left(  
  \frac{d^3F_{\gamma}}{d\varpi^3} 
+ 2\frac{d^2 F_{\gamma}}{d\varpi^2} \left(2F_{\gamma} +1\right) 
+ \frac{dF_{\gamma}}{d\varpi} 
- 2 \left(\frac{d F_{\gamma}}{d\varpi}\right)^2 
\right)  
\Bigg].
\nonumber  
\end{align}
The first line above is the usual Kompaneets result \cite{Kompaneets:1957}. 
The following two lines are the lowest order relativistic corrections in $T/m_{\chi} = 1/x$ obtained by keeping higher-order terms in the transfer energy expansion~\cite{Challinor_1998x}.  Note that the elastic collision term vanishes for the equilibrium distribution with a chemical potential $F_{\gamma} (\varpi,x) = 1/[A(x) e^{\varpi}-1]$, with $A(x)$ being any function of $x$.

The approximation $\omega \ll m_\chi$ may not seem satisfactory for the dark photons produced in DM annihilations and bound-state decays with energies $\omega \sim m_{\chi}$. As we shall see however, for these high-energy dark photons, the thermalisation rate estimated according to the above exceeds the rate for any other process by tens of orders of magnitude (cf.~\cref{fig:Rates_AnnDec}). Our conclusions will thus be robust, and the approximation is sufficient for our purposes.

\subsubsection{Bound-state formation and ionisation \label{sec:PhotonDist_CollisionTerms_BSF}}

The ionisation contribution to the collision term \eqref{eq:Collision_BSF} can be expressed as 
\begin{align}
C_{\cal B}^{\rm BSF} \supset
- \omega \,f_{\gamma}(\omega,t) 
\, \Gamma^{\mathsmaller{\gamma {\cal B} \rightarrow \chi\bar{\chi}}} (\omega) ,
\label{eq:C_}
\end{align}
where $\Gamma^{\gamma {\cal B} \rightarrow \chi\bar{\chi}} (\omega,t)$ is the rate at which a dark photon of energy $\omega$ ionises bound states of type ${\cal B}$,
\begin{align}
\Gamma^{\gamma {\cal B} \rightarrow \chi\bar{\chi}} (\omega,t) \equiv
g_{\cal B} \int \frac{d^3 p_{\cal B}}{(2\pi)^3} f_{\cal B}(E_{\cal B}) 
(\sigma_{\cal B}^{\rm ion} v_{\mathsmaller{\text{M\o{}l}}}) ,
\label{eq:_GammaIon}
\end{align}
with $v_{\mathsmaller{\text{M\o{}l}}}=1$ being the $\gamma {\cal B}$ M\o{}ller velocity. 
Note that the rate \eqref{eq:_GammaIon}, which is proportional to the density of the bound states, should be discerned from the ionisation rate of a bound state ${\cal B}$, $\Gamma^{\rm ion}_{\cal B}$,  defined in \cref{eq:GammaIonB_def}, which is proportional to the density of the ionising radiation. 
From \cref{eq:omega,eq:Milne} it follows that in the non-relativistic regime,  $\sigma_{\cal B}^{\rm ion}$ depends at leading order only on $\omega$ and not on $p_{\cal B}$. With this, \cref{eq:_GammaIon} becomes
\begin{align}
\label{eq:_GammaIon_final}
\Gamma^{\gamma {\cal B} \rightarrow \chi \bar{\chi}} (\omega, t) 
\simeq n_{\cal B} \sigma_{\cal B}^{\rm ion}.
\end{align}
By taking into account the principle of detailed balance and that the BSF contribution should be proportional to $n_{\chi}^2 [1+f_{\gamma}(\omega,t)]$, we find that the collision term \eqref{eq:Collision_BSF} is
\begin{align}
C_{\cal B}^{\rm BSF} \simeq
- \omega \left[f_{\gamma}(\omega,t) n_{\cal B} 
- n_{\cal B}^{\rm eq} 
\, f_{\gamma}^{\rm eq}(\omega)
\, \frac{1+f_{\gamma}(\omega,t)}{1+f_{\gamma}^{\rm eq}(\omega,t)} 
\, \frac{n_{\chi}^2}{(n_{\chi}^{\rm eq})^2}\right]
\sigma_{\cal B}^{\rm ion} .
\label{eq:C__NR}
\end{align}

Note that for the dark photons produced in the $\chi \bar{\chi}$ annihilation or bound-state decay, $\omega \sim m_{\chi}$ and $v_{\rm rel} \rightarrow 1$, thus the relations \cref{eq:omega,eq:Milne} as well as the existing computations of $\sigma_{\cal B}^{\mathsmaller{\rm BSF}}$ become inaccurate. However, it is easy to see that $\sigma_{\cal B}^{\rm ion}$ decreases quickly with increasing $\omega$. The high-energy dark photons thus do not contribute to $C_{\cal B}^{\rm BSF}$ significantly before they get redshifted down to low energies due to the expansion of the universe and thermalisation (cf.~\cref{fig:Rates_AnnDec}). We may therefore use the same formulae without introducing significant inaccuracies to the calculation.

\subsubsection{Bound-to-bound transitions \label{sec:PhotonDist_CollisionTerms_B2B}}

The excitation contribution to the collision term \eqref{eq:Collision_B2B} is
\begin{align}
C_{{\cal B}{\cal B}\prime}^{\rm B2B} \supset 
-\omega f_{\gamma}(\omega,t) 
\, \Gamma^{\gamma {\cal B}\rightarrow{\cal B}'}  (\omega,t) ,
\end{align}
where $\Gamma^{\gamma{\cal B}\rightarrow{\cal B}\prime}$ is the rate at which a dark photon upscatters bound states ${\cal B}$ to ${\cal B}'$,
\begin{align}
\Gamma^{\gamma{\cal B}\rightarrow{\cal B}'}  (\omega,t)
\equiv  
g_{\cal B} \int \frac{d^3p_{\cal B}}{(2\pi)^3} 
\, f_{\cal B}(E_{\cal B}) 
\, (\sigma_{\gamma{\cal B}\rightarrow{\cal B}'} \, v_{\mathsmaller{\text{M\o{}l} }}).
\label{eq:B2B_GammaUpscattering_def}
\end{align}
Here, $v_{\mathsmaller{\text{M\o{}l} }} = G_{\gamma{\cal B}}/(\omega E_{\cal B})$ is the $\gamma {\cal B}$ M\o{}ller velocity, with 
$G_{\gamma{\cal B}} = p_{\gamma}\cdot p_{\cal B} = (\mathbb{s}-m_{\cal B}^2)/2$ and $\mathbb{s}$ being the first Mandelstam variable (cf.~\cref{App:CrossSections}). The excitation cross-section $\sigma_{\gamma{\cal B}\rightarrow{\cal B}'}$ is related to the de-excitation rate $\Gamma_{{\cal B}'\rightarrow\gamma{\cal B}}^0$ at the ${\cal B}'$ rest frame as follows [cf.~\cref{eq:DecayAndInverseDecay}]
\begin{align}
\sigma_{\gamma{\cal B}\rightarrow{\cal B}'} v_{\mathsmaller{\text{M\o{}l}}}
&= 
\Gamma_{{\cal B}'\rightarrow\gamma{\cal B}}^0 \times
4\pi^2
\, \frac{g_{{\cal B}'}}{g_{\gamma} g_{\cal B}}
\, \frac{m_{{\cal B}'}^2}{\omega E_{\cal B}(m_{{\cal B}'}^2 - m_{\cal B}^2)}
\delta(\sqrt{\mathbb{s}} - m_{{\cal B}'}) 
\nonumber \\
&= 
\Gamma_{{\cal B}' \rightarrow\gamma{\cal B}}^{0} \times
4\pi^2
\, \frac{g_{{\cal B}'}}{g_{\gamma} g_{\cal B}}
\, \frac{m_{{\cal B}'}^3}{\omega^2 p_{\cal B} E_{\cal B} (m_{{\cal B}'}^2 - m_{\cal B}^2)}
\, \delta \left[\cos \theta_{\gamma{\cal B}} -\frac{1}{p_{\cal B}}\left(E_{\cal B} - \frac{m_{{\cal B}'}^{2} - m_{\cal B}^{2}}{2\omega} \right) \right].
\end{align}
With this, \cref{eq:B2B_GammaUpscattering_def} becomes
\begin{align}
\Gamma^{\gamma{\cal B}\rightarrow{\cal B}'} =
\Gamma_{{\cal B}'\rightarrow\gamma{\cal B}}^0
\, \frac{g_{{\cal B}'}}{g_{\gamma}}
\, \frac{m_{{\cal B}'}^3}{\omega^2 (m_{{\cal B}'}^2 - m_{\cal B}^2)}
\, \int_{m_{\cal B}}^{\infty} d E_{\cal B} f_{\cal B} (E_{\cal B}) 
\, \Theta \left(
1-\left|\frac{1}{p_{\cal B}}\left(E_{\cal B} - \frac{m_{{\cal B}'}^2 - m_{\cal B}^2}{2\omega} \right)\right|
\right).
\label{eq:B2B_GammaUpscattering_step}
\end{align}
We find that the requirement $|\cos \theta_{\gamma{\cal B}}| \leqslant 1$ imposes
\begin{align}
E_{\cal B} \geqslant \frac{m_{\cal B}^2 + E_c(\omega)^2}{2E_c(\omega)}
\qquad \text{with} \qquad
E_c (\omega) \equiv \frac{m_{{\cal B}'}^2-m_{\cal B}^2}{2\omega}.
\end{align}
This condition ensures also that $E_{\cal B} \geqslant m_{\cal B}$. 
Assuming further that the bound states are in kinetic equilibrium,
their occupation numbers are
\begin{align}
f_{\cal B} (E_{\cal B}) = e^{-(E_{\cal B}-\mu_{\cal B})/T} = (n_{\cal B}/n_{\cal B}^{\rm eq}) e^{-E_{\cal B}/T},
\label{eq:fB_KineticEquil}
\end{align}
with $\mu_{\cal B}$ being their chemical potential. Then, \cref{eq:B2B_GammaUpscattering_step} becomes
\begin{align}
\Gamma^{\gamma{\cal B}\rightarrow{\cal B}'} =
\Gamma_{{\cal B}'\rightarrow\gamma{\cal B}}^0
\, \frac{n_{\cal B}}{n_{\cal B}^{\rm eq}} 
\, \frac{g_{{\cal B}'}}{g_{\gamma}}
\, \frac{m_{{\cal B}'}^3 T}{\omega^2 (m_{{\cal B}'}^2 - m_{\cal B}^2)}
\, \exp \left[-\frac{m_{\cal B}^2 + E_c(\omega)^2}{2E_c(\omega) T} \, \right] .
\label{eq:B2B_GammaUpscattering}
\end{align}
Collecting everything, and taking into account the principle of detailed balance and that the de-excitation contribution, 
${\cal B}' \rightarrow {\cal B} + \gamma$,  
should be proportional to $[1+f_{\gamma}(\omega,t)]n_{{\cal B}'}$, we find that the collision term \eqref{eq:Collision_BSF} due to bound-to-bound transitions is
\begin{align}
C_{{\cal B}{\cal B}'}^{\rm B2B} &= - \omega 
\, \Gamma_{{\cal B}{\cal B}'}^{\rm B2B}
\left(
f_{\gamma}(\omega,t) \frac{n_{\cal B}}{n_{\cal B}^{\rm eq}} 
- f_{\gamma}^{\rm eq}(\omega,t) \frac{1+f_{\gamma}(\omega,t)}{1+f_{\gamma}^{\rm eq}(\omega,t)} \frac{n_{{\cal B}'}}{n_{{\cal B}'}^{\rm eq}} \right) ,
\label{eq:CB2B}
\end{align}
where we defined 
\begin{align}
\Gamma_{{\cal B}{\cal B}'}^{\rm B2B} &\equiv
\Gamma_{{\cal B}'\rightarrow\gamma{\cal B}}^0
\, \frac{g_{{\cal B}'}}{g_{\gamma}}
\, \frac{m_{{\cal B}'}^3 T}{\omega^2 (m_{{\cal B}'}^2 - m_{\cal B}^2)}
\, \exp \left[- \frac{\omega}{T} \left(\frac{m_{\cal B}^2}{m_{{\cal B}'}^2 - m_{\cal B}^2}\right) - \frac{m_{{\cal B}'}^2-m_{\cal B}^2}{4\omega T} \right] .
\label{eq:B2B_GammaB2B}
\end{align}
At temperatures $T \ll m_{\cal B}$, which encompasses the range of interest for the purpose of estimating the radiation backreaction, the exponential factor above implies that $C_{{\cal B}{\cal B}'}^{\rm B2B}$ is suppressed for all $\omega$, even those on resonance, $\omega \approx \Delta {\cal E}_{\cal B}$.

\subsubsection{Bound-state decays and inverse decays \label{sec:PhotonDist_CollisionTerms_BSD}}

For simplicity, we will consider here only the (inverse) decays of the bound states into two dark photons. The (inverse) decays into three dark photons are not expected to change our conclusions. 

The inverse decay contribution to the collision term \eqref{eq:Collision_BSD2} is
\begin{align}
C_{\cal B}^{\rm BSD} \supset -\omega f_{\gamma}(\omega,t) 
\, \Gamma_{\cal B}^{\rm BSD}  (\omega,t)
\end{align}
with
\begin{align}
\Gamma_{\cal B}^{\rm BSD}  (\omega,t)
\equiv  
g_{\gamma} \int \frac{d^3p_{\gamma}'}{(2\pi)^3} 
\, f_{\gamma}(\omega',t) 
\, \sigma_{\gamma\gamma\rightarrow{\cal B}} \, (1-\cos \theta_{\gamma\gamma}).
\label{eq:BSD_GammaInverseDecay_def}
\end{align}
with $\cos \theta_{\gamma\gamma} = \hat{\bf p}_{\gamma} \cdot \hat{\bf p}_{\gamma}'$. The inverse decay cross-section 
$\sigma_{\gamma\gamma\rightarrow{\cal B}}$ 
is related to the decay rate 
$\Gamma_{{\cal B}\rightarrow\gamma\gamma}^0$ 
at the ${\cal B}'$ rest frame as follows [cf.~\cref{eq:DecayAndInverseDecay}]
\begin{align}
\sigma_{\gamma\gamma\rightarrow{\cal B}} = 
\Gamma_{{\cal B}\rightarrow\gamma\gamma}^0 \times
\, \frac{g_{\cal B}}{g_{\gamma}^2}
\, \frac{8\pi^2}{m_{\cal B}^2}
\delta(\sqrt{s} - m_{\cal B}) 
\end{align}
or equivalently
\begin{align}
\sigma_{\gamma\gamma\rightarrow{\cal B}}  \, (1-\cos \theta_{\gamma\gamma})
&= 
\Gamma_{{\cal B}\rightarrow\gamma\gamma}^0 \times
4\pi^2 \, \frac{g_{\cal B}}{g_{\gamma}^2}
\, \frac{m_{\cal B}}{\omega^2\omega'^2}
\, \delta \left[\cos \theta_{\gamma\gamma} - \left(1-\frac{m_{\cal B}^2}{2\omega\omega'}\right) \right] ,
\end{align}
With this, \cref{eq:BSD_GammaInverseDecay_def} becomes
\begin{align}
\Gamma_{\cal B}^{\rm BSD} =
\Gamma_{{\cal B}\rightarrow\gamma\gamma}^0
\, \frac{g_{\cal B}}{g_{\gamma}}
\, \frac{m_{\cal B}}{\omega^2}
\, \int_{m_{\cal B}^2/(4\omega)}^{\infty} d\omega' f_{\gamma} (\omega',t)  .
\label{eq:BSD_GammaInverseDecay}
\end{align}
Note that the lower limit on $\omega'$ is found from the requirement $|\cos \theta_{\gamma{\cal B}}| \leqslant 1$. For an equilibrium dark photon distribution, $f_{\gamma}^{\rm eq}(\omega,t) \simeq e^{-\omega/T}$, 
\begin{align}
\Gamma_{\cal B}^{\rm BSD, eq} =
\Gamma_{{\cal B}\rightarrow\gamma\gamma}^0
\, \frac{g_{\cal B}}{g_{\gamma}}
\, \frac{m_{\cal B} T }{\omega^2}
\, e^{-m_{\cal B}^2/(4\omega T)}  
=
\Gamma_{{\cal B}\rightarrow\gamma\gamma}^0
\, \frac{g_{\cal B}}{g_{\gamma}}
\, \frac{x}{\varpi^2}
\, e^{-x^2/(4\varpi)} .
\label{eq:BSD_GammaInverseDecay_Equil}
\end{align}
Collecting everything, and taking into account the principle of detailed balance and that the decay contribution, ${\cal B} \rightarrow \gamma\gamma$, should be proportional to $n_{\cal B}$, we find that the collision term \eqref{eq:Collision_BSD2} due to bound-state decays and inverse decays is
\begin{align}
C_{\cal B}^{\rm BSD} &= - \omega
\left(f_{\gamma}(\omega,t) \, \Gamma_{\cal B}^{\rm BSD}
- f_{\gamma}^{\rm eq}(\omega,t) \, \Gamma_{\cal B}^{\rm BSD, eq} \, \frac{n_{\cal B}}{n_{\cal B}^{\rm eq}} \right) .
\label{eq:CBSD}
\end{align}

\subsubsection{Annihilation and pair creation \label{sec:PhotonDist_CollisionTerms_Ann}}

The pair creation contribution to the collision term \eqref{eq:Collision_Ann} is
\begin{align}
C^{\rm ann}\supset-\omega f_{\gamma} (\omega,t) \, \Gamma^{\rm pair} (\omega,t) ,
\label{eq:Cann_1}
\end{align}
where 
\begin{align}
\Gamma^{{\rm pair}} (\omega,t) \equiv
g_{\gamma}\int \frac{d^3p_{\gamma}'}{(2\pi)^3} f_{\gamma}(\omega',t) 
\, \sigma^{\gamma\gamma\rightarrow X\bar{X}}
\, (1-\cos \theta_{\gamma\gamma}) ,
\label{eq:Ann_GammaPair_def}
\end{align}
with $\cos \theta_{\gamma\gamma} = \hat{\bf p}_{\gamma} \cdot \hat{\bf p}_{\gamma}'$.\footnote{Note that the factor $(1-\cos \theta_{\gamma\gamma})$ in \cref{eq:Ann_GammaPair_def} is the $\gamma\gamma$ M\o{}ller velocity, which is in general \emph{not} equal to 1. Omitting this factor leads to incorrect results; upon averaging, it results in a factor 1/2.}  
The pair-creation cross-section is related to the annihilation cross-section using \cref{eq:CrossSections_InverseProcesses}
\begin{align}
\sigma^{\gamma\gamma\rightarrow \chi\bar{\chi}} = 
\sigma^{\chi\bar{\chi} \rightarrow\gamma\gamma} 
\, \frac{g_{\chi}^2}{g_{\gamma}^2}
\, \left(\frac{G_{\chi \Bar{\chi}}}{G_{\gamma \gamma}}\right)^2 ,
\end{align}
where $G_{\gamma \gamma}$ and $G_{\chi \Bar{\chi}}$ are defined as in \cref{eq:Gfactor_def}. 
In the non-relativistic regime, $G_{\chi \Bar{\chi}}\simeq m_{\chi}^2 v_{\rm rel}$, and from \cref{eq:Gfactors_Inverse} follows that $G_{\gamma \gamma} = m_{\chi}^2+\sqrt{G_{\chi \Bar{\chi}}^2+m_{\chi}^4}$. Using the above, we may express $\sigma^{\gamma\gamma\rightarrow \chi\bar{\chi}}$ in terms of $v_{\rm rel}$ and exchange the integration over $\cos\theta_{\gamma\gamma}$ in \cref{eq:Cann_1} for integration over $v_{\rm rel}$. Conservation of energy imposes 
\begin{align}
-1\leqslant \cos\theta_{\gamma\gamma} \leqslant 1- \frac{2m_{\chi}^2}{\omega\omega'}
\qquad \text{or} \qquad
0 \leqslant v_{\rm rel} \leqslant v_{\max} = 
2\sqrt{\frac{\omega\omega'}{m_{\chi}^2} \left(\frac{\omega\omega'}{m_{\chi}^2}-1\right) } .
\end{align}
and $\omega\omega'/m_{\chi}^2 \geqslant 1$.
With this, we find
\begin{align}
\Gamma^{{\rm pair}} (\omega,t) =
\frac{g_{\chi}^2}{4\pi^2 g_{\gamma}} 
\int_{m_{\chi}^2/\omega}^\infty d\omega' \omega'^2 f_{\gamma}(\omega',t) 
\, \tau(\omega\omega'/m_{\chi}^2)
\label{eq:GammaPair}
\end{align}
where
\begin{align}
\tau (b) 
&= \frac{1}{b^2}
\int_0^{2\sqrt{b(b-1)}} d v_{\rm rel}
\, \frac{v_{\rm rel}^3}{\left(1+\sqrt{1+v_{\rm rel}^2}\right) \sqrt{1+v_{\rm rel}^2}}
\, \sigma^{\chi \bar{\chi} \rightarrow \gamma\gamma}
\nonumber \\ &
\simeq
\frac{1}{2b^2}
\int_0^{2\sqrt{b(b-1)}} d v_{\rm rel} \, v_{\rm rel}^2
(\sigma^{\chi \bar{\chi} \rightarrow \gamma\gamma} v_{\rm rel}) ,
\label{eq:Cann_tau}
\end{align}
with $b\geqslant 1$, and in the last step we kept only the leading order term, consistently with our non-relativistic approximation. Putting \cref{eq:GammaPair,eq:Cann_tau} together, we may inverse the order of integration,
\begin{subequations}
\label{eq:GammaPair2}
\label[pluralequation]{eqs:GammaPair2}
\begin{align}
\Gamma^{{\rm pair}} (\omega,t) \simeq
\frac{g_{\chi}^2}{g_{\gamma}} 
\frac{m_{\chi}^4}{8\pi^2 \omega^2} 
\int_0^{\infty} d v_{\rm rel} \, v_{\rm rel}^2
(\sigma^{\chi \bar{\chi} \rightarrow \gamma\gamma} v_{\rm rel})
\int_{m_{\chi}^2 (1+v_{\rm rel}^2/4)/\omega}^\infty d\omega'
f_{\gamma}(\omega',t) .
\label{eq:Ann_GammaPair_NonConf}
\end{align}
\end{subequations}
For the equilibrium dark photon density, \cref{eq:Ann_GammaPair_NonConf} becomes
\begin{align}
\Gamma^{{\rm pair},{\rm eq}} (\omega,t)
&\simeq 
\frac{g_{\chi}^2}{4g_{\gamma}} 
\left(\frac{m_{\chi}^2 T^5}{\pi^3 \omega}\right)^{1/2} 
e^{-m_{\chi}^2/(\omega T)}
\nonumber \\
&\times \left[
\frac{[m_{\chi}^2/(\omega T)]^{3/2}}{2\sqrt{\pi}}
\int_0^{\infty} d v_{\rm rel} \, v_{\rm rel}^2
\, (\sigma^{\chi \bar{\chi} \rightarrow \gamma\gamma} v_{\rm rel})
e^{-[m_{\chi}^2/(\omega T)] v_{\rm rel}^2/4} \right],
\label{eq:Ann_GammaPair_NonConf_Equil}
\end{align}
where the factor in the square brackets corresponds to the thermally averaged annihilation cross-section \eqref{eq:SEannthermallyavegared} at effective temperature $T_{\rm eff} = \omega T/m_\chi$.

Considering the principle of detailed balance and that the annihilation contribution to the collision term \eqref{eq:Collision_Ann} must be proportional to $n_{\chi}^2$, we find the full expression
\begin{equation}
  C^{\rm ann} = -\omega \left[
f_{\gamma} (\omega,t) \, \Gamma^{{\rm pair}}(\omega,t) - 
f_{\gamma}^{\rm eq} (\omega,t) \, \Gamma^{{\rm pair},{\rm eq}}(\omega,t) 
\left(\frac{n_{\chi}}{n_{\chi}^{\rm eq}}\right)^2
\right].
\label{eq:Cann}  
\end{equation}

\subsection{Boltzmann equation for dark-photon distribution \label{sec:PhotonDist_BoltzmannEq_Final}}

Collecting together the Liouville operator \eqref{eqs:LiouvilleOp_final} and the collision terms computed in \cref{sec:PhotonDist_CollisionTerms}, the Boltzmann equation for the dark photon energy distribution becomes
\begin{align}
\label{eq:BoltzmannEq_final}
\frac{\partial F_{\gamma}(\varpi,x)}{\partial x} =
&+\frac{2\lambda}{x^3\varpi^2} \frac{\partial}{\partial\varpi} 
\left[ \varpi^4 j(\varpi,x) \right]  Y_{\chi} \sigma_{T} 
\nonumber \\
&-\frac{\lambda}{x^2}  
\sum_{\cal B} 
\left[\frac{F_{\gamma}(\varpi,x)}{F_{\gamma}^{\rm eq}(\varpi)}  \frac{Y_{\cal B}}{Y_{\cal B}^{\rm eq}} - \frac{1+F_{\gamma}(\varpi,x)}{1+F_{\gamma}^{\rm eq}(\varpi,x)}
\frac{Y_{\chi}^2}{(Y_{\chi}^{\rm eq})^2} \right] 
F_{\gamma}^{\rm eq}(\varpi) \, Y_{\cal B}^{\rm eq}(x) \, \sigma_{\cal B}^{\rm ion} (\varpi,x)
\nonumber \\ 
&-\Lambda \, x \sum_{{\cal B},{\cal B}'}
\left[\frac{F_{\gamma}(\varpi,x)}{F_{\gamma}^{\rm eq}(\varpi)}  
\frac{Y_{\cal B}}{Y_{\cal B}^{\rm eq}}
- \frac{1+F_{\gamma}(\varpi,x)}{1+F_{\gamma}^{\rm eq}(\varpi,x)}
\frac{Y_{{\cal B}'}}{Y_{{\cal B}'}^{\rm eq}} \right]  
F_{\gamma}^{\rm eq}(\varpi)
\, \Gamma_{{\cal B}{\cal B}'}^{\rm B2B} (\varpi,x)
\nonumber \\ 
&-\Lambda \, x \sum_{\cal B}
\left[\frac{F_{\gamma}(\varpi,x)}{F_{\gamma}^{\rm eq}(\varpi)}  
\frac{\Gamma_{\cal B}^{\rm BSD} (\varpi,x)}{\Gamma_{\cal B}^{{\rm BSD},{\rm eq}} (\varpi,x)} 
- \frac{Y_{\cal B}}{Y_{\cal B}^{\rm eq}} \right] 
F_{\gamma}^{\rm eq}(\varpi) \, \Gamma_{\cal B}^{{\rm BSD, eq}} (\varpi,x)
\nonumber \\ 
&-\Lambda \, x
\left[\frac{F_{\gamma}(\varpi,x)}{F_{\gamma}^{\rm eq}(\varpi)}  
\frac{\Gamma^{{\rm pair}} (\varpi,x)}{\Gamma^{{\rm pair},{\rm eq}} (\varpi,x)} 
- \frac{Y_{\chi}^2}{(Y_{\chi}^{\rm eq})^2} \right] 
F_{\gamma}^{\rm eq}(\varpi) \, \Gamma^{{\rm pair},{\rm eq}} (\varpi,x).
\end{align}
Note that the pair-creation and inverse-decay terms in the above render this equation integro-differential due to the $\Gamma^{{\rm pair}}$ and $\Gamma_{\cal B}^{\rm BSD}$ factors.

\subsection{Dark-photon interaction rates \label{sec:PhotonDist_Rates}}

The Boltzmann \cref{eq:BoltzmannEq_final} for the phase-space distribution of the dark photons could be solved using the results from \cref{Sec:FreezeOut} for $Y_{\chi}(x)$ and $Y_{\cal B}(x)$ in order to investigate whether $F_\gamma(\varpi,x)$ deviates significantly from the equilibrium value assumed in that computation. Considering however the challenge of solving a partial integro-differential equation, we begin by comparing the rates of the different processes involved. 
As we shall see, this suffices to show robustly that the backreaction of the radiation produced during the DM thermal decoupling is negligible. 

\subsubsection{Definition of rates \label{sec:PhotonDist_Rates_def}}

For each process, we compute the contribution to the logarithmic derivative
\begin{align}
\left| \dfrac{x}{F_{\gamma}}
\dfrac{dF_{\gamma}(\varpi,x)}{dx} \right|_{\rm process} =
\dfrac{R^{\rm process}}{H} ,
\label{eq:dlnFdlnx}
\end{align}
where $R^{\rm process} \equiv f_{\gamma}(\omega,t)^{-1} [\partial f_{\gamma} (\omega,t) / \partial t]_{\rm process}$ is the relative rate of change of the dark-photon occupation number due a specific process. 
The normalised rate \eqref{eq:dlnFdlnx} being larger than 1 implies that a perturbation of order 1 in the dark photon spectrum, $\delta F_{\gamma} \sim F_\gamma$, can be potentially smoothed out via this interaction within one Hubble time, $\delta x \sim x$. Since we are concerned with the interactions in which a dark photon produced during the DM thermal decoupling may participate, we consider the elastic scattering, bound-state ionisation, excitation, inverse decay and DM pair-creation rates, rather than the rates of the inverse processes.

We use the density $Y_{\chi}(x)$ found by the freeze-out computation of \cref{Sec:FreezeOut} and $Y_{\cal B}(x)$ as estimated via the generalised Saha equilibrium \cref{eq:SahaGen}. For all processes but the elastic scattering, we compute \cref{eq:dlnFdlnx} at the equilibrium dark-photon distribution. In this way, we consider the following
\begin{subequations}
\label{eq:timescales}
\begin{align}
\dfrac{R^{\rm elas}}{H} 
&\approx 
\dfrac{2 \lambda}{x^2}
\dfrac{Y_{\chi}}{F_{\gamma}}
\dfrac{1}{\varpi^2} 
\left|
\dfrac{\partial}{\partial\varpi} 
\left[ \varpi^4 j(\varpi,x) \right] 
\right|
\, \sigma_{T} ,
\label{eq:rate_elas}
\\
\dfrac{R_{\cal B}^{\rm ion}}{H} 
&\approx 
\dfrac{\lambda}{x} 
\left(\dfrac{Y_{\chi}}{Y_{\chi}^{\rm eq}}\right)^2 
Y_{\cal B}^{\rm eq}(1-r_{\cal B})
\, \sigma_{\cal B}^{\rm ion} ,
\label{eq:rate_BSI}
\\
\dfrac{R_{\cal B \to B'}^{\rm excit}}{H} 
&\approx 
x^2 \Lambda 
\left(\dfrac{Y_{\chi}}{Y_{\chi}^{\rm eq}}\right)^2 
(1-r_{\cal B})
\, \Gamma_{{\cal B}{\cal B}'}^{\rm B2B},
\label{eq:rate_B2B}
\\
\dfrac{R_{\cal B}^{\rm inv.dec}}{H} 
&\approx 
x^2 \Lambda 
\, \Gamma_{\cal B}^{{\rm BSD, eq}},
\label{eq:rate_BSID}
\\
\dfrac{R^{\rm pair}}{H} 
&\approx 
x^2 \Lambda 
\, \Gamma^{\rm pair, eq}.
\label{eq:rate_pair}
\end{align}
\end{subequations}

The elastic term \eqref{eq:rate_elas} merits some more discussion. In contrast to all other terms, it includes the inverse processes, i.e.~both Compton and inverse Compton scattering. By virtue of detailed balance, it thus vanishes for $F_{\gamma} = F_{\gamma}^{\rm eq}$. In order to calculate the equilibration rate via elastic scattering and compare it with the rates of the other processes, we must consider a non-equilibrium $F_{\gamma}$. It is convenient to parametrise the departure from equilibrium by assuming that the dark photons have an equilibrium-like distribution at a temperature different than that of DM, $T_{\gamma} \neq T$; this sets $F_{\gamma} = (e^{a\varpi}-1)^{-1}$,  where $a\equiv T_{\gamma} / T$. 

\begin{figure}[t!]
\centering
\includegraphics[width=0.98\linewidth]{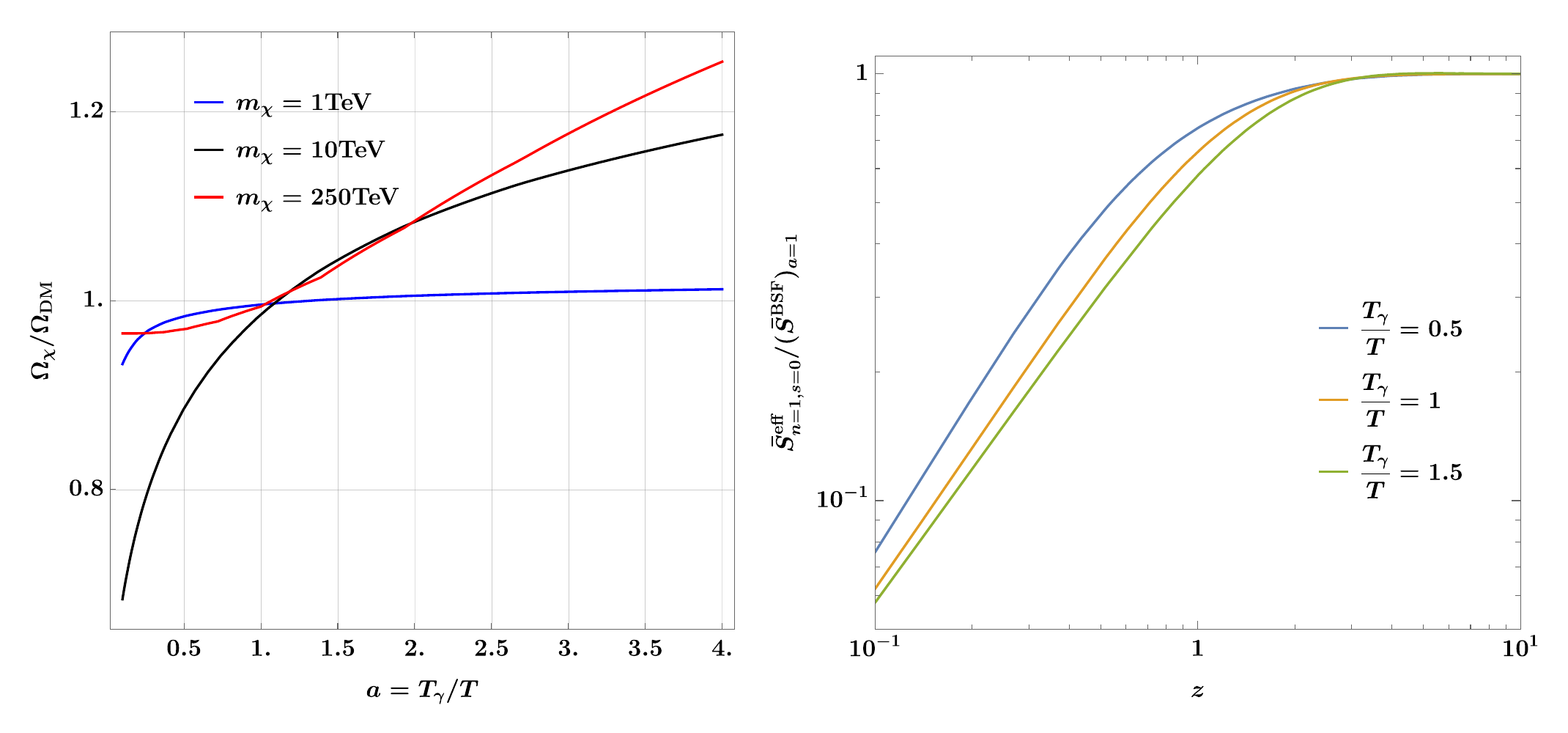}
\caption{
\emph{Left:} Ratio of the $\chi$ relic density to the observed DM density as a function of the dark-photon-to-DM temperature ratio, $a = T_\gamma/ T$, for different values of the DM mass and the corresponding values of the coupling $\alpha (m_{\chi})$ that yield the observed DM abundance for $a=1$. $a$ affects the bound-state ionisation rate (see plot on the right), and consequently $\Omega_{\chi}$. 
\emph{Right:} Increasing (decreasing) the dark-photon temperature enhances (suppresses) the bound-state ionisation processes, thereby suppressing (enhancing) the effective DM depletion rate via BSF.   We show the contribution to the effective DM depletion cross-section due to $\chi\bar{\chi}$ capture to the spin-singlet ground state, normalised to the corresponding thermally averaged BSF cross-section for $a = 1$, as a function of $z=m_{\chi} \alpha^2 / (4T)$.}
\label{fig:achoice}
\end{figure}

We choose $a$ as follows. If $a$ is too close to 1, the elastic scattering rate is very small and would appear subdominant to all other rates; however, in such a case, thermalisation is already (nearly) achieved, and there is no radiation backreaction. (The effect of the thermal radiation on the bound states is already included in the freeze-out computations.) In the opposite limit, if $a$ is very different from 1, the elastic scattering rate becomes very large, which may lead to overestimating the rate of this process with respect to those of other processes. To be conservative, we thus choose the minimum $a>1$ that, if sustained, would yield a significant change on the DM relic density with respect to the $a=1$ case. Considering the $\sim 1\%$ experimental uncertainty in $\Omega_{\rm DM}$, we deem a significant change in the computed relic density to be 10\%.  The impact of $a$ on $\Omega_{\chi}$ is shown in \cref{fig:achoice}.

\subsubsection{Relevant cosmological epoch \label{sec:PhotonDist_Rates_Epoch}}

We are interested in comparing the dark-photon interaction rates in the epoch during which the DM depletion via BSF is significant.  As already discussed, this begins typically when the plasma temperature approaches the ground-state binding energy; it is then that the ionisation rate(s) become lower than the decay rates of the bound states and allow for the effective DM depletion cross-section to saturate to the actual one. The exact time depends on the details of model, in the present model being when 
$|{\cal E}_{n=1}|/T \gtrsim 0.28$, with $|{\cal E}_{n=1}| = m_{\chi} \alpha^2/4$~\cite{vonHarling:2014kha}. However, if this occurs before freeze-out, we only need to consider temperatures $T \lesssim T_{\rm fo} \approx m_{\chi} / 30$, as everything is determined by the thermal equilibrium before that. For every set of parameters, we also determine the temperature at which the DM density reaches within 10\% from the final relic density (determined by integrating the Boltzmann equation of \cref{Sec:FreezeOut} until very late times); we refer to this as the temperature of the DM thermal decoupling, $T_{\rm th.dec.}$, and emphasise that it is much lower than the freeze-out temperature; $m_{\chi}/T_{\rm th.dec.} \sim 10^3 - 10^4$ in the relevant mass range. In summary, we will consider roughly the range 
$\max(30, 1/\alpha^2) \lesssim  m_{\chi}/T \lesssim 10^4$.

\subsubsection{Dark-photon injection energies, and comparison of interaction rates   \label{sec:PhotonDist_Rates_Energies}}

We shall consider dark photons produced in BSF and bound-to-bound transitions, whose energies are around the bound-state binding energies and the splittings between them, as well as dark photons produced in DM direct annihilations or bound-state decays with energies close to the DM mass. In each case we compare the pertinent interaction rates.

\subparagraph{Resonant dark photons from bound-to-bound transitions.} 

\begin{figure}[t!]
\centering    
\includegraphics[width=0.98\linewidth]{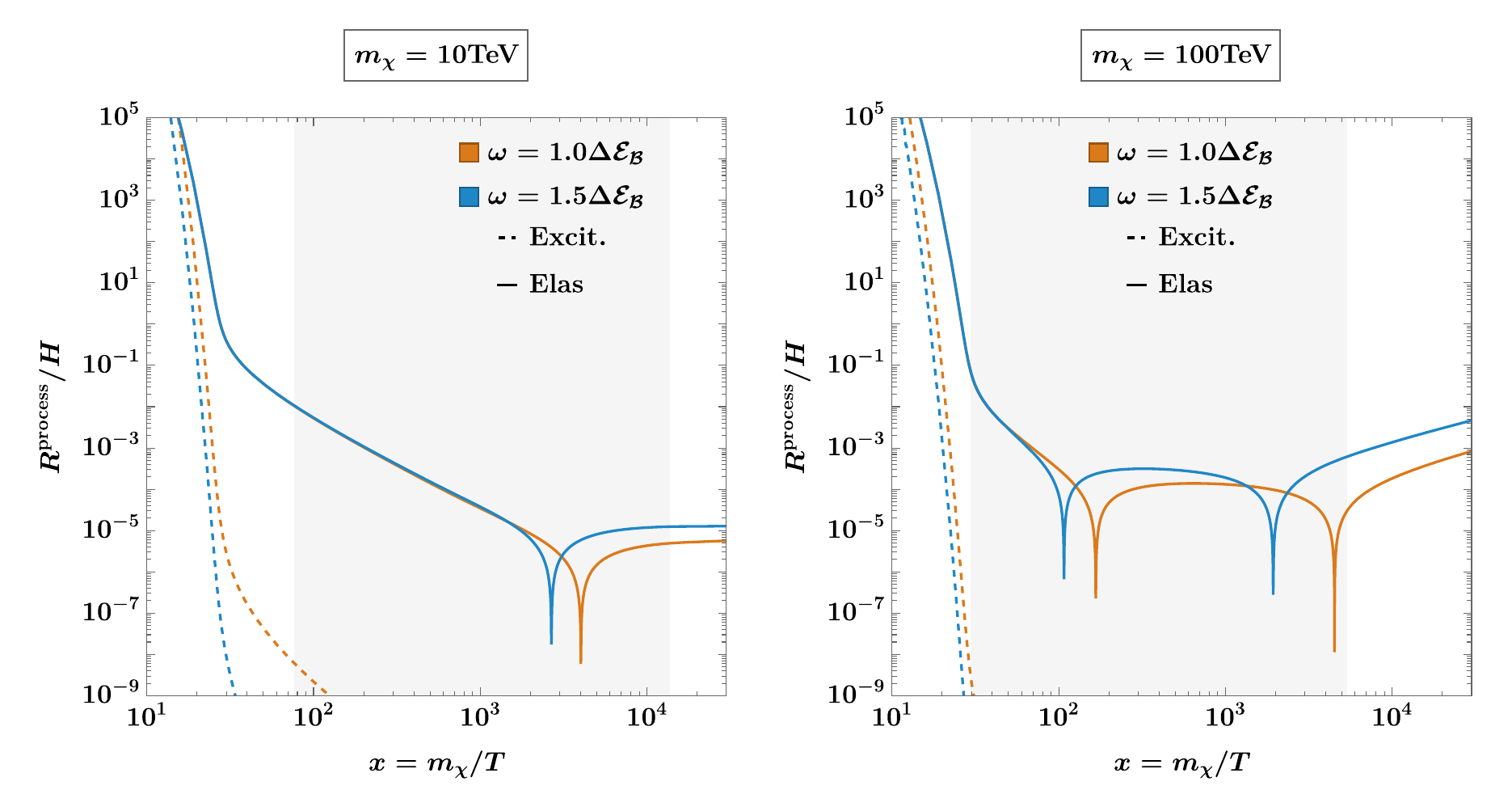}
\caption{The bound-state excitation and elastic scattering rates normalised to the Hubble rate for dark photons of energy $\omega$ close to the energy splitting between the first excited and the ground state. 
The resonant dark photons are much more likely to redshift out of resonance, rather than scatter elastically with free DM particles, and more likely to scatter elastically with DM and thermalise than excite bound states.
The spikes in the elastic scattering rate correspond to a change in sign in the  corresponding collision term \eqref{eq:Collision_elas}. The shaded region indicates the epoch after freeze-out within which the DM depletion via BSF is significant (cf.~\cref{sec:PhotonDist_Rates_Epoch}).}
\label{fig:Rates_B2B}
\end{figure}

The dark photons produced in de-excitation processes have energies around 
$\Delta {\cal E}_{\cal B} = {\cal E}_{n=2}-{\cal E}_{n=1} = 3m_{\chi} \alpha^2/16$ with a thermal broadening due the bound-state velocities that can be up to 
$|\delta \omega| 
\lesssim \sqrt{T/ M_{\cal B}}~\Delta {\cal E}_{\cal B}
\lesssim \alpha \Delta {\cal E}_{\cal B} < \Delta {\cal E}_{\cal B}$, for  the DM masses of interest.
In \cref{fig:Rates_B2B}, we compare the rates at which such dark photons
scatter elastically on free DM particles, excite bound states or redshift due to the Hubble expansion, for the indicative DM masses of 10 TeV and 100 TeV. The elastic scattering rate overwhelmingly dominates over the bound-state excitation rate, while they both remain several orders of magnitude lower than the Hubble rate that redshifts these dark photons out of resonance. 
We have considered $\omega = \Delta {\cal E}_{\cal B}$ and $1.5 \Delta {\cal E}_{\cal B}$ (the result for $\omega = 0.5 \Delta {\cal E}_{\cal B}$ is similar to the latter), noting that the excitation rate becomes even more negligible for photon energies outside this range. This results from the exponential suppression appearing in \cref{eq:B2B_GammaB2B}.

\subparagraph{Resonant dark photons from bound-state formation.} 

\begin{figure}[t!]
\centering
\includegraphics[width=0.98\linewidth]{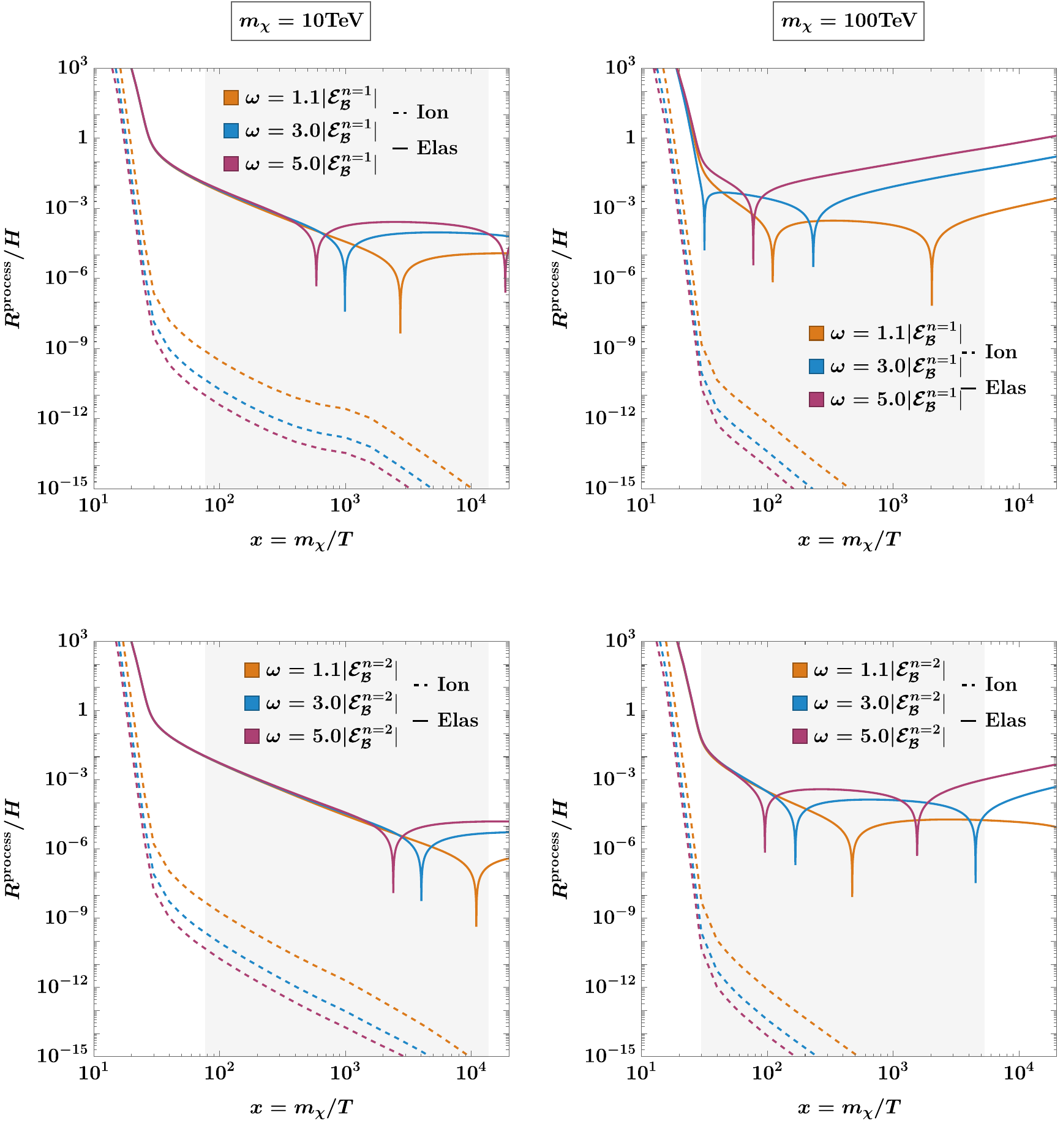}
\caption{ Rates for thermalisation via elastic scattering and ionisation of a bound state to the Hubble rate for a dark photon of energy $\omega$ close to the binding energy of the bound level considering  the ground state only (\textit{top}) and the excited state only (\textit{bottom}) for two indicative masses $m_{\chi} = 10, 100$ TeV with the required coupling to reproduce DM density. These low-energy dark photons are more likely to redshift rather than scatter off of DM particles or ionise bound states. The shaded region indicates the epoch after freeze-out within which the DM depletion via BSF is significant (cf.~\cref{sec:PhotonDist_Rates_Epoch}).}
\label{fig:Rates_BSF}
\end{figure}

The dark photons produced in capture processes have energies close to the binding energies ${\cal E}_{\cal B}$ of the corresponding bound levels, augmented by $\delta\omega \sim T$ due to the thermal velocities of the free particles~[cf.~\cref{eq:omega}]; evidently $\delta \omega \lesssim \omega$ in the range of interest. The bound-state ionisation and elastic scattering rates are shown in \cref{fig:Rates_BSF}. The elastic scattering rate is several orders of magnitude larger than the ionisation rate during the relevant times. The ratio of the ionisation to the elastic scattering rate decreases with time according to $R^{\rm ion}/  R^{{\rm elas}} \propto (Y_{\cal B} \sigma^{\rm BSF})/(Y_{\chi} \sigma^{{\rm elas}})$, where $Y_{\mathcal{B}} \ll Y_{\chi}$ as deduced from the generalised Saha equilibrium \cref{eq:SahaGen} (cf.~\cref{sec:FreezeOut_GenSahaEquil}).

Moreover, these rates remain consistently several orders of magnitude lower than the Hubble rate. It is thus much more likely that the dark photons produced in BSF processes redshift out of resonance rather than scatter with free particles. Scattering with free DM particles and thermalising is in turn much more likely than the dark photons ionising bound states. These observations hold true across the entire relevant dark-photon energy range, and become more pronounced for larger dark-photon energies, since the bound-state ionisation cross-section decreases with increasing dark photon energy. 

Finally we note that for the resonant dark photons produced either in bound-to-bound transition or BSF processes, pair creation and bound-state inverse decays are entirely negligible, since such low-energy dark photons need a partner dark photon of much higher energy in order for pair creation and inverse bound-state decay to be energetically possible; this introduces the exponential suppression seen in \cref{eq:Ann_GammaPair_NonConf_Equil,eq:BSD_GammaInverseDecay_Equil}.

\subparagraph{High-energy dark photons from DM annihilations and bound-state decays.}  

\begin{figure}[t!]
\centering
\includegraphics[width=0.98\linewidth] {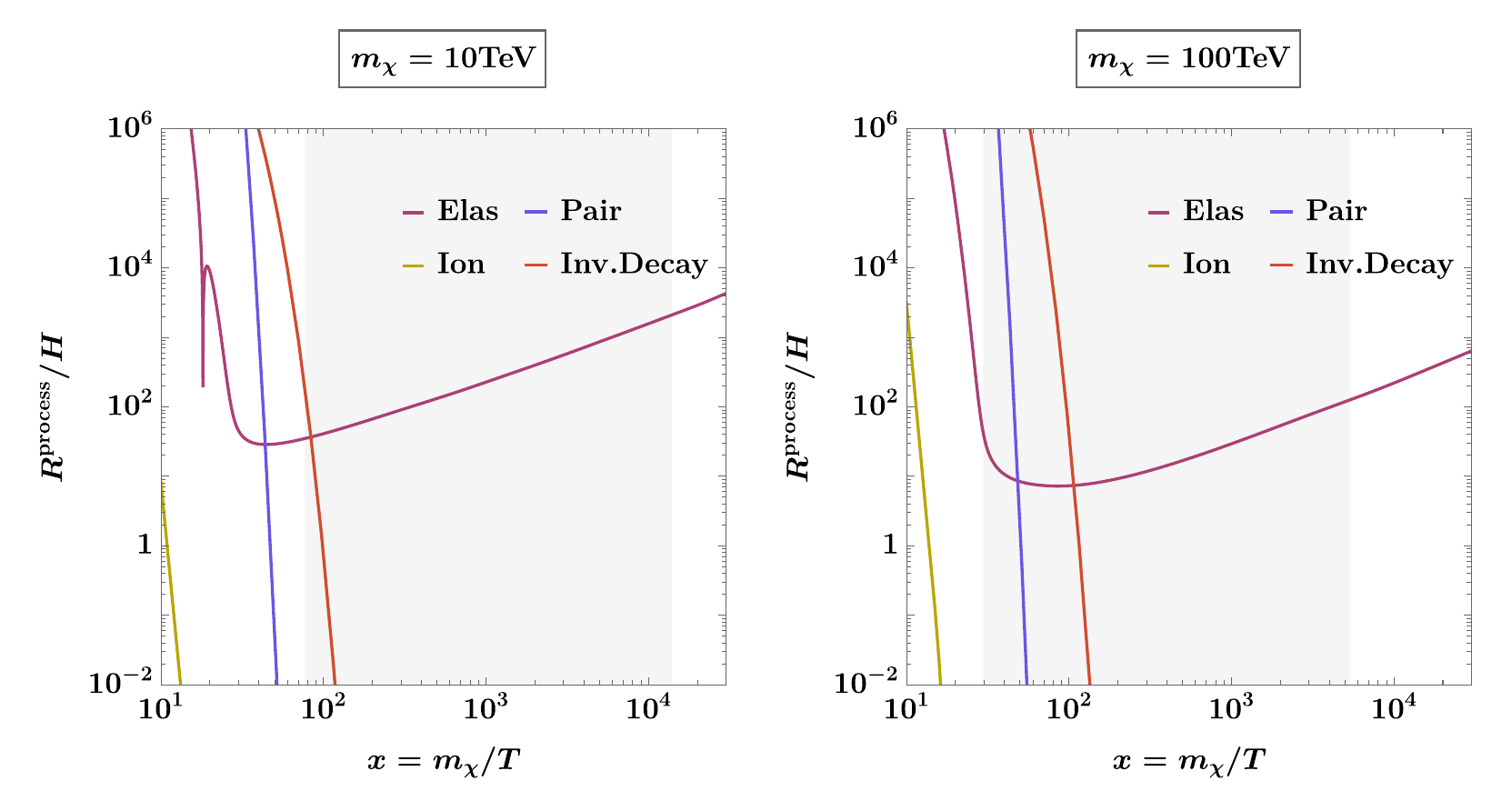}
\caption{The pair-creation, inverse bound-state decay, elastic scattering and ionisation rates normalised to the Hubble rate for a dark photon of energy $\omega = m_{\chi}$ as a function of $x$, and for the two indicative DM masses $m_{\chi} = 10$ and $100$~TeV with the required coupling to reproduce DM density. The high-energy dark photons are more likely scatter off of unbound DM particles and thermalise, rather than redshift, ionise bound states, pair-create DM particles or inverse-decay into bound states. The shaded region indicates the epoch after freeze-out within which the DM depletion via BSF is significant (cf.~\cref{sec:PhotonDist_Rates_Epoch}).}
\label{fig:Rates_AnnDec}
\end{figure}

In \cref{fig:Rates_AnnDec} we compare the interaction rates for dark photons produced with energy $\omega = m_{\chi}$. 
At early times, before BSF effects are important, the inverse decay and pair creation can be faster than both elastic scattering and redshifting due to the Hubble expansion, as expected from the DM freeze-out calculation. However, at lower temperatures, when BSF effects become significant, elastic scattering dominates. This suggests that the high-energy dark photons produced in annihilation and bound-state decay processes are more likely to thermalise, rather than redshift, pair-create DM particles, inverse-decay into bound states or ionise bound states.  
In fact, the difference between the thermalisation rate and the rate of any other process spans orders of magnitude, ensuring that the above conclusion is robust despite the fact that the approximations for the elastic scattering and ionisation rates we have used are not fully reliable at such high dark-photon energies. 

Dark photons produced with $\omega \sim m_{\chi}$ that escape prompt thermalisation (or other interactions) and redshift down to lower energies, will interact according to the probabilities shown in \cref{fig:Rates_B2B,fig:Rates_BSF}. That is, when redshifted into the potentially dangerous resonant energy range, such dark photons are more likely to continue redshifting, rather than participating in any other process.

\subsection{Other models \label{sec:PhotonDist_OtherModels}}

The results presented in \cref{sec:PhotonDist_Rates} demonstrate in the context of the model adopted here that the back-reaction of the dark photons produced during the DM thermal decoupling is negligible. These dark photons are overwhelmingly more likely to thermalise or redshift out of resonance than participate in any process that impedes DM depletion: ionisation, excitation, inverse decay of bound states, or pair-creation of free DM particles. 

We expect these conclusions to carry out, and in fact strengthen, in other models. In particular, the metastability of the bound states that can deplete the DM density ensures that $Y_{\cal B} / Y_{\rm DM}$ is always suppressed, as follows from the generalised Saha equilibrium \cref{eq:SahaGen} discussed in \cref{sec:FreezeOut_GenSahaEquil}. This in turn implies that the ratio of bound-state ionisation or excitation to elastic scattering rates will be suppressed. 

If the theory contains additional light dofs that couple to the ionising radiation, then the thermalisation of the radiation produced during the DM thermal decoupling occurs even faster. This may be the case in less-minimal $U(1)$ models with darkly-charged particles lighter than DM, or in similar scenarios with scalar mediators.  It is also the case in non-Abelian theories where the gauge bosons can thermalise via self-scattering.

Furthermore, many models, including Abelian~\cite{Oncala:2019yvj} and non-Abelian~\cite{Harz:2018csl} theories, feature repulsive long-range interactions in the scattering states that participate in BSF. The Sommerfeld effect then suppresses exponentially the corresponding BSF cross-sections at low velocities, and, by virtue of the Milne relation, the ionisation cross-sections close to resonance. 

\section{Conclusion \label{Sec:Conclusions}}

The emergence of a new type of inelasticity -- the formation of bound states -- is a feature of viable thermal-relic DM scenarios around and above the TeV scale. This is supported by model-independent unitarity arguments~\cite{Baldes:2017gzw} and confirmed by model-specific computations~\cite{vonHarling:2014kha}. Metastable bound states introduce DM annihilation channels that can dominate the DM depletion rate in the early universe. Incorporating these channels in the computation of the DM thermal decoupling, and doing so accurately, is therefore essential for predicting correctly the DM parameters and consequently the expected experimental signals. The latter is particularly important in view of existing and upcoming multi-TeV cosmic-ray observatories, such as H.E.S.S., C.T.A. and IceCube, that are or will be probing annihilating DM in this mass range. 

To compute the effect of metastable bound states on the DM abundance accurately, we must take into account the bound-state dissociation processes, which suppress the DM depletion via such channels. Up to now, calculations have considered only the ionisation of bound states by the radiation of the thermal bath. However, the paradigm of Hydrogen formation in the early universe suggests that the radiation produced in the formation of bound states may back-react and re-ionise matter very efficiently, having exactly the right frequency to do so.
In the case of DM decoupling, the DM direct annihilations and the decay of the metastable DM bound states provide additional sources of radiation that could potentially ionise or excite bound states, even though this radiation has energies of the order of the DM mass, i.e.~far from the resonance range.

This work provides the first study in the literature of the possibility that the radiation produced during the DM thermal decoupling, resonant and not, may back-react and hinder the DM depletion via metastable bound states. For this purpose, we have formulated the Boltzmann equation for the phase-space density of the dark radiation. Leveraging results from DM freeze-out calculations and using the generalised Saha equilibrium equation for the densities of metastable bound states, we have shown that the low-energy resonant dark radiation produced in BSF and bound-to-bound transitions is more likely to redshift before ionising or exciting a bound state, while the higher-energy dark radiation produced in DM annihilations or bound-state decays is more likely to thermalise via elastic scattering on free DM particles rather than engage into any other process. The difference between competing rates reaches even orders of magnitude in the relevant parameter space. 

The difference with respect to Hydrogen recombination arises mostly from the suppressed density of the metastable bound states, as predicted by the generalised Saha equilibrium~\cite{Binder:2021vfo}. In the case of stable bound states, such as Hydrogen, the low entropy of the plasma renders bound states the energetically favourable configuration at temperatures below the binding energy; the probability that the resonant radiation backreacts is large because the density of bound states becomes significant.
In the case of metastable bound states, the bound-state decay ensures that their densities remain always low, even as they provide a very efficient channel for DM depletion; the radiation backreaction is thus negligible because there are very few bound states present to be ionised. 

While we have carried out our computations in the context of a minimal gauged $U(1)$ model, our conclusions are only expected to strengthen in more complex models, where there are typically additional light species off of which dark radiation can scatter and thermalise even faster. Our study thus shows robustly that the radiation backreaction can be neglected in the computations of the DM thermal decoupling. This is important for essentially all models that feature DM at the TeV scale and above produced via thermal decoupling from the primordial plasma. 
\appendix

\section*{Appendix}

\section{Cross-sections: some useful formulae \label{App:CrossSections}}
We consider the 2-to-$n$ process
\begin{align}
A (k_{\mathsmaller{A}}) + B (k_{\mathsmaller{B}}) \rightarrow 
1 (p_1)  + \cdots + n  (p_n) ,
\label{eq:Process_2-to-n}
\end{align}
where the parentheses denote the four-momenta for each particle. As standard, we define the Mandelstam variables
\begin{align}
\mathbb{s} \equiv (k_{\mathsmaller{A}} + k_{\mathsmaller{B}})^2
\quad \text{and} \quad
\mathbb{t}_j \equiv (k_{\mathsmaller{A}} - p_{j})^2,
\label{eq:Mandelstam}
\end{align}
where for 2-to-2 scattering, $\mathbb{t}_1$ and $\mathbb{t}_2$ are simply denoted by $\mathbb{t}$ and $\mathbb{u}$.

In the following, $|{\mathcal M}_{\mathsmaller{A+B \leftrightarrow 1+ \cdots + n}}|^2$ denotes the squared amplitude summed over the dofs of the initial and final state, such as spin, gauge dofs or others.

\subsubsection*{Total cross-section}
The cross-section for the process \eqref{eq:Process_2-to-n} is
\begin{align}
\sigma_{\mathsmaller{A+B \rightarrow 1+ \cdots + n}} = 
\frac{1}{g_{\mathsmaller{A}}g_{\mathsmaller{B}}}
\frac{1}{4G_{\mathsmaller{AB}}} \int d\Pi_1 \cdots \int d\Pi_n 
|\mathcal{M}_{\mathsmaller{A+B \leftrightarrow 1+ \cdots + n}}|^2
(2\pi)^4 \delta^4 (k_{\mathsmaller{A}}+k_{\mathsmaller{B}} - p_1 - \cdots - p_n),
\label{eq:CrossSection_def}
\end{align}
where the M\o{}ller flux factors $G_{\mathsmaller{AB}}$ are
\begin{align}
G_{\mathsmaller{AB}} \equiv 
\sqrt{ (p_{\mathsmaller{A}} \cdot p_{\mathsmaller{B}})^2 - m_{\mathsmaller{A}}^2 m_{\mathsmaller{B}}^2 }
= \sqrt{ (\mathbb{s}-m_{\mathsmaller{A}}^2-m_{\mathsmaller{B}}^2)^2/4 - m_{\mathsmaller{A}}^2 m_{\mathsmaller{B}}^2 } 
=E_A E_B \, v_{\mathsmaller{\text{M\o{}l}}},
\label{eq:Gfactor_def}
\end{align}
with $v_{\mathsmaller{\text{M\o{}l}}}$ being the $AB$ M\o{}ller velocity.

\subsubsection*{2-to-2 scattering}
For 2-to-2 scattering, the differential cross-section is
\begin{align}
\frac{d\sigma_{\mathsmaller{A+B \rightarrow 1+2}}}{d\mathbb{t}} = 
\frac{1}{g_{\mathsmaller{A}}g_{\mathsmaller{B}}}
\frac{1}{4G_{\mathsmaller{AB}}^2}
\frac{|\mathcal{M}_{\mathsmaller{A+B \leftrightarrow 1+2}}|^2}{16 \pi}
\label{eq:DifferentialCrossSection_2to2}
\end{align}
where in terms of the CM variables, 
$d\mathbb{t} = 2 |{\bf k}_{\mathsmaller{A}}^{\rm CM}| |{\bf p}_1^{\rm CM}| d\cos \theta^{\rm CM}$, with  
$|{\bf k}_{\mathsmaller{A}}^{\rm CM}| = G_{\mathsmaller{AB}} / \sqrt{\mathbb{s}}$ and
$|{\bf p}_{\mathsmaller{1}}^{\rm CM}| = G_{\mathsmaller{12}} / \sqrt{\mathbb{s}}$ 
($G_{\mathsmaller{12}}$ is defined analogously to \cref{eq:Gfactor_def} for the pair of particles $1,2$).

Assuming no CP violation, the amplitudes for inverse processes are equal. Then, considering \cref{eq:DifferentialCrossSection_2to2}, the cross-sections for two inverse 2-to-2 processes are related as follows
\begin{align}
\frac
{\sigma_{\mathsmaller{A+B \rightarrow 1+2}}}
{\sigma_{\mathsmaller{1+2 \rightarrow A+B}}} = 
\frac{g_1 g_2}{g_{\mathsmaller{A}}g_{\mathsmaller{B}}}
\left(\frac{G_{\mathsmaller{12}}}{G_{\mathsmaller{AB}}}\right)^2 .
\label{eq:CrossSections_InverseProcesses}
\end{align}
where $G_{\mathsmaller{AB}}$ and $G_{\mathsmaller{12}}$ are related as follows
\begin{align}
\mathbb{s}=
2\sqrt{G_{\mathsmaller{AB}}^2 + m_{\mathsmaller{A}}^2m_{\mathsmaller{B}}^2} + m_{\mathsmaller{A}}^2 + m_{\mathsmaller{B}}^2 =
2\sqrt{G_{\mathsmaller{12}}^2 + m_{\mathsmaller{1}}^2m_{\mathsmaller{2}}^2} + m_{\mathsmaller{1}}^2 + m_{\mathsmaller{2}}^2.
\label{eq:Gfactors_Inverse}
\end{align}

\subsubsection*{Decays and inverse decays}
For $A+B \leftrightarrow \chi$, the inverse-decay cross-section and decay rate (in the $\chi$ rest frame) are
\begin{align}
\sigma_{\mathsmaller{A+B} \rightarrow \chi} &= 
\frac{1}{g_{\mathsmaller{A}} g_{\mathsmaller{B}}}
\frac{1}{4G_{\mathsmaller{AB}}} 
\frac{|\mathcal{M}_{\mathsmaller{A+B} \leftrightarrow \chi}|^2}{2m_{\chi}}
(2\pi) \delta(\sqrt{\mathbb{s}}-m_{\chi}) ,
\label{eq:InverseDecayCrossSection}
\\
\Gamma_{\chi \rightarrow \mathsmaller{A+B}}^0 &= \frac{1}{g_{\chi}} \frac{G_{\mathsmaller{AB}}}{8\pi m_{\chi}^3} 
\, |\mathcal{M}_{\mathsmaller{A+B} \leftrightarrow \chi}|^2 ,
\label{eq:DecayRate}
\end{align}
where we took into account that $|p_{\mathsmaller{A}}^{\rm CM}|=|p_{\mathsmaller{B}}^{\rm CM}|=G_{\mathsmaller{AB}}/ \sqrt{\mathbb{s}}$.
From the above, we deduce that
\begin{align}
\frac{\sigma_{\mathsmaller{A+B} \rightarrow \chi}}{\Gamma_{\chi \rightarrow \mathsmaller{A+B}}^0} = \frac{g_{\chi}}{g_{\mathsmaller{A}}g_{\mathsmaller{B}}}
\, \frac{\pi m_{\chi}^2}{G_{\mathsmaller{AB}}^2}
\, (2\pi) \delta(\sqrt{\mathbb{s}}-m_{\chi}) .
\label{eq:DecayAndInverseDecay}
\end{align}

\acknowledgments

We thank Tobias Binder for useful discussions. This work was supported by the European Union’s Horizon 2020 research and innovation programme under grant agreement No 101002846, ERC CoG CosmoChart.

\bibliography{Bibliography}

\end{document}